\documentclass[journal=jacsat,manuscript=article]{achemso}

\usepackage{titlesec}
\usepackage{chemformula} 
\usepackage[T1]{fontenc} 
\usepackage{amsfonts}
\usepackage{multirow}
\usepackage{tabulary}
\usepackage{IEEEtrantools}

\usepackage{amsmath,siunitx,booktabs,graphicx}
\usepackage{hyperref}
\usepackage[capitalize]{cleveref}

\author{Haoyu Lin}
\affiliation[CCME]{BNLMS, State Key Laboratory for Structural Chemistry of Unstable \& Stable Species, College of Chemistry and Molecular Engineering, Peking University, Beijing 100871, China}
\alsoaffiliation[CCSE]{Center for Computational Science and Engineering, Peking University, Beijing 100871, China}
\author{Shiwei Wang}
\affiliation[CCME]{BNLMS, State Key Laboratory for Structural Chemistry of Unstable \& Stable Species, College of Chemistry and Molecular Engineering, Peking University, Beijing 100871, China}
\author{Jintao Zhu}
\affiliation[CQB]{Center for Quantitative Biology, Academy for Advanced Interdisciplinary Studies, Peking University, Beijing 100871, China}
\author{Yibo Li}
\affiliation[CLS]{Center for Life Sciences, Academy for Advanced Interdisciplinary Studies, Peking University, Beijing 100871, China}
\author{Jianfeng Pei}
\email{jfpei@pku.edu.cn}
\affiliation[CQB]{Center for Quantitative Biology, Academy for Advanced Interdisciplinary Studies, Peking University, Beijing 100871, China}
\author{Luhua Lai}
\email{lhlai@pku.edu.cn}
\affiliation[CCME]{BNLMS, State Key Laboratory for Structural Chemistry of Unstable \& Stable Species, College of Chemistry and Molecular Engineering, Peking University, Beijing 100871, China}
\alsoaffiliation[CCSE]{Center for Computational Science and Engineering, Peking University, Beijing 100871, China}
\alsoaffiliation[CLS]{Center for Life Sciences, Academy for Advanced Interdisciplinary Studies, Peking University, Beijing 100871, China}
\alsoaffiliation[CQB]{Center for Quantitative Biology, Academy for Advanced Interdisciplinary Studies, Peking University, Beijing 100871, China}

\title[DeepRLI]{DeepRLI: A Multi-objective Framework for Universal Protein--Ligand Interaction Prediction}

\abbreviations{IR,NMR,UV}
\keywords{Protein, Ligand, Binding Affinity, Graph Transformer, Contrastive Learning}

\begin{document}







\begin{abstract}
  Protein (receptor)--ligand interaction prediction is a critical component in computer-aided drug design, significantly influencing molecular docking and virtual screening processes. Despite the development of numerous scoring functions in recent years, particularly those employing machine learning, accurately and efficiently predicting binding affinities for protein--ligand complexes remains a formidable challenge. Most contemporary methods are tailored for specific tasks, such as binding affinity prediction, binding pose prediction, or virtual screening, often failing to encompass all aspects, indicating a lack of comprehensive understanding of free energy and limitations in generalization. In this study, we put forward DeepRLI, a novel protein--ligand interaction prediction architecture. It encodes each protein--ligand complex into a fully connected graph, retaining the integrity of the topological and spatial structure, and leverages the improved graph transformer layers with cosine envelope as the central module of the neural network, thus exhibiting superior scoring power. In order to equip the model to generalize to conformations beyond the confines of crystal structures and to adapt to molecular docking and virtual screening tasks, we propose a multi-objective strategy, that is, the model outputs three scores for scoring and ranking, docking, and screening, and the training process optimizes these three objectives simultaneously. For the latter two objectives, we augment the dataset through a docking procedure, incorporate suitable physics-informed blocks and employ an effective contrastive learning approach. Eventually, our model manifests a balanced performance across scoring, ranking, docking, and screening, thereby demonstrating its ability to handle a range of tasks. Overall, this research contributes a multi-objective framework for universal protein--ligand interaction prediction, augmenting the landscape of structure-based drug design.
\end{abstract}

\section{Introduction}

Drug discovery aims to identify active molecules, namely lead compounds, capable of binding to disease-related biological targets \cite{blass2021basic}. In the realm of drug design, the development of a scoring function that can accurately quantifying the interaction between a receptor and a ligand facilitates the discovery of lead compounds via computer-assisted techniques \cite{gore2018computational,singh2020computer}. Theoretically, a perfect scoring function corresponds to the free energy surface. Utilizing this, we can employ a geometric optimizer to find the minimum point of the free energy landscape, thus realizing molecular docking, and obtaining the stable binding pose of the receptor and the ligand. Subsequently, the calculation of their binding free energy, commonly referred to as binding affinity, can be conducted. And based on this, a further step is able to carry out virtual screening, allowing the identification of small molecular ligands with the strongest binding to the receptor, which are potential candidate drugs \cite{blass2021basic,gore2018computational,singh2020computer}.

However, the potential energy landscape of receptor--ligand systems is highly complex, and even with various approximations, the evaluation of free energy based on the principles of statistical mechanics is still computationally intensive and time-consuming \cite{pathria2021statistical,chipot2007free}. Scoring functions are actually born for high-throughput virtual screening, needing to strike an optimal balance between speed and accuracy. So, in essence, these functions serve as significantly simplified approximations for free energy estimation. Their methodology eschews dependence on the state density in phase space, electing instead to derive rough estimates of free energy directly from a single conformation \cite{Baron2012computational}. A range of traditional scoring functions, including physics-based, empirical, and knowledge-based approaches, have shown basic proficiency in meeting these requirements and have been widely used in docking and screening tasks \cite{liu2015classification,li2019anoverview}.

Recent years have witnessed an exponential increase in experimental and computed structure data of large biomolecules \cite{berman2000theprotein,burley2022rcsb} and substantial advancements in artificial intelligence algorithms \cite{dara2022machine,cerchia2023new}. This concurrent progression has sparked considerable research interest in machine learning methods, which offer superior expressiveness and obviate the need for manual rule-setting, to develop better scoring functions. Numerous machine learning-based scoring functions that take 3D structures as input have already emerged \cite{meli2022scoring}. These functions typically excel in part of areas, \textit{e.g.}, binding affinity prediction, binding pose prediction, or virtual screening, significantly outperforming the traditional counterparts. For instance, based on crystal structures, methodologies like $K_{\text{DEEP}}$ \cite{jimenez2018kdeep} and InteractionGraphNet \cite{jiang2021interactiongraphnet} are capable of inferring affinity scores with a high linear correlation to experimental binding data. Additionally, techniques such as DeepDock \cite{mendezlucio2021ageometric} and RTMScore \cite{shen2022boosting} demonstrate impressive capability to accurately discern native binding poses from a pool of computer-generated decoy conformations and efficiently identify the true binders within a collection of decoy molecules for a specified target receptor. Nevertheless, very few machine learning-based scoring functions have shown consistently outstanding performance across all tasks, indicating a need for continued research to optimize these models for comprehensive applicability.

An ideal scoring function should exhibit excellent performance across all key metrics, including scoring power, ranking power, docking power, and screening power. The task-specificity of some methods points towards their limited generalizability. This phenomenon is primarily attributable to the biased nature of the training data and the absence of inherent physical insights in machine learning algorithms, thereby posing challenges for models in making accurate inferences from unseen data. As a response, numerous data augmentation strategies have been put forth \cite{francoeur2020three,morrone2020combining,stafford2022atomnet}. However, most past strategies have predominantly sacrificed the prediction of binding affinity values, instead in favor of classification models which offer broader practical application \cite{lim2019predicting}. This shift arises primarily because augmented data does not provide concrete free energy values. Besides, there have been attempts to amalgamate traditional scoring functions with machine learning to enhance conventional methods. These endeavors involve introducing energy correction terms into classical equations \cite{wang2017improving,lu2019incorporating,yang2022delta} and leveraging latent space representations to parameterize physics-inspired formulas \cite{moon2022pignet}. Furthermore, a recent development is the GenScore model proposed by Shen \textit{et al.}, which achieves a balanced multi-task performance by correlating machine learning's statistical potentials with experimental binding data \cite{shen2023ageneralized}. Notably, those methods striving for multi-aspect performance all involve the profile of traditional scoring functions to some extent.

In this work, we propose DeepRLI, a novel deep learning model for protein--ligand interaction prediction. It adopts an innovative multi-objective strategy, namely outputting multiple scores simultaneously to suit various different tasks, thereby realizing a versatile and universally applicable machine learning scoring function that demonstrates a balanced and exceptional ability in scoring, ranking, docking, and screening. Specifically, our model adopts an improved graph transformer with cosine envelope as its principal feature embedding module to obtain the hidden representation of each atom. Subsequently, three independent readout modules output the scoring score, docking score, and screening score respectively. Among them, the scoring score is used for the binding free energy prediction task of the protein--ligand complex crystal structure, suitable for lead compound optimization scenarios; the docking score is instrumental in ascertaining the most favorable binding posture between a protein and a ligand; the screening score is utilized to assess the potency of various small molecules against designated targets.

Theoretically, a model that can accurately predict the free energy difference between any state of any receptor--ligand structure and its dissociation state would be considered an ideal and powerful scoring function. However, it encounters practical problems due to the scarcity of available data. The existing complex structure data, along with their associated binding free energy information, is markedly limited, which poses a significant challenge for the development of deep learning scoring functions that rely exclusively on data-driven approaches to precisely estimate the relative free energies of various receptor--ligand conformations. In light of this, we incorporate reasonable physics-informed blocks into both the docking readout module and the screening readout module, thus ensuring the model's generalization ability. Additionally, we expand the training data by re-docking and cross-docking crystal structure data employing a molecular docking program. Based on the fact that the native binding conformation is located at the minimum point of the free energy surface, and the free energy of other conformations must inherently exceed it, we conceive an effective contrastive learning method to optimize parameters, which enables the model to grasp the relation between the free energy values of different structures.

Overall, our protein--ligand interaction scoring model, DeepRLI, achieves exceptional versatility and efficacy across a range of tasks through a divide-and-conquer multi-objective approach combined with data augmentation and contrastive learning strategy, reaching the state-of-the-art level in scoring, ranking, docking, and screening. Moreover, it is worth mentioning that the inherent attention mechanism and physics-inspired constraint blocks in the model endow it with excellent interpretability. The atom pairs assigned higher attention weights and larger physical scores by the model correspond to key interactions, such as hydrophobic interactions, hydrogen bonds, and $\pi$-stacking. This indicates that our universal scoring model accurately captures information related to interactions, thereby exhibiting outstanding performance.

\section{Methods}

DeepRLI is a novel deep learning-based scoring function specifically designed for predicting protein--ligand interactions. It employs a sophisticated graph neural network architecture to accurately evaluate the binding strength of the specific 3D complex structures. The underlying methodology and detailed framework of DeepRLI are elucidated in the subsequent sections.

Graph neural networks, first proposed by Scarselli, Gori \textit{et al.} \cite{scarselli2004graphical,gori2005new,scarselli2009graph}, are deep learning models specifically tailored for the graph domains. In recent years, GNNs have gained significant attention as a widely employed graph analysis method, primarily due to their exceptional performance \cite{liu2020introduction,wu2022graph}. The crucial component of GNNs is the graph convolution operation, which serves as a generalization of function convolution. Through a series of approximations and simplifications, the graph convolution expression commonly employed in machine learning is the version proposed by Kipf and Welling \cite{kipf2017semisupervised}. From a spatial perspective, this can be interpreted as a message passing (also known as neighborhood aggregation) paradigm \cite{gilmer2017neural}. Messages are generated based on the neighboring environment, and the representations of nodes and edges are updated in accordance with these messages.

In the field of chemistry, GNNs possess four advantages over traditional 3D-CNNs: First, the graph topology ensures the rotation and reflection invariance of the prediction results. Second, the neural network can accept graphs with any number of nodes as input, allowing for the complete inclusion of molecules of any size without padding or truncation. Third, there is no need for space voxelization as in 3D-CNNs, thus avoiding potential instability caused by a large number of empty grid points. Fourth, distance information can be accurately embedded in the graph, whereas the distance accuracy of 3D-CNNs depends on the interval of grid points. Consequently, utilizing graphs to represent molecules, as the method adopted here, for the prediction of molecular properties is highly suitable.

\subsection{Architecture}

\begin{figure}[!hbtp]
  \includegraphics[height=0.75\textheight]{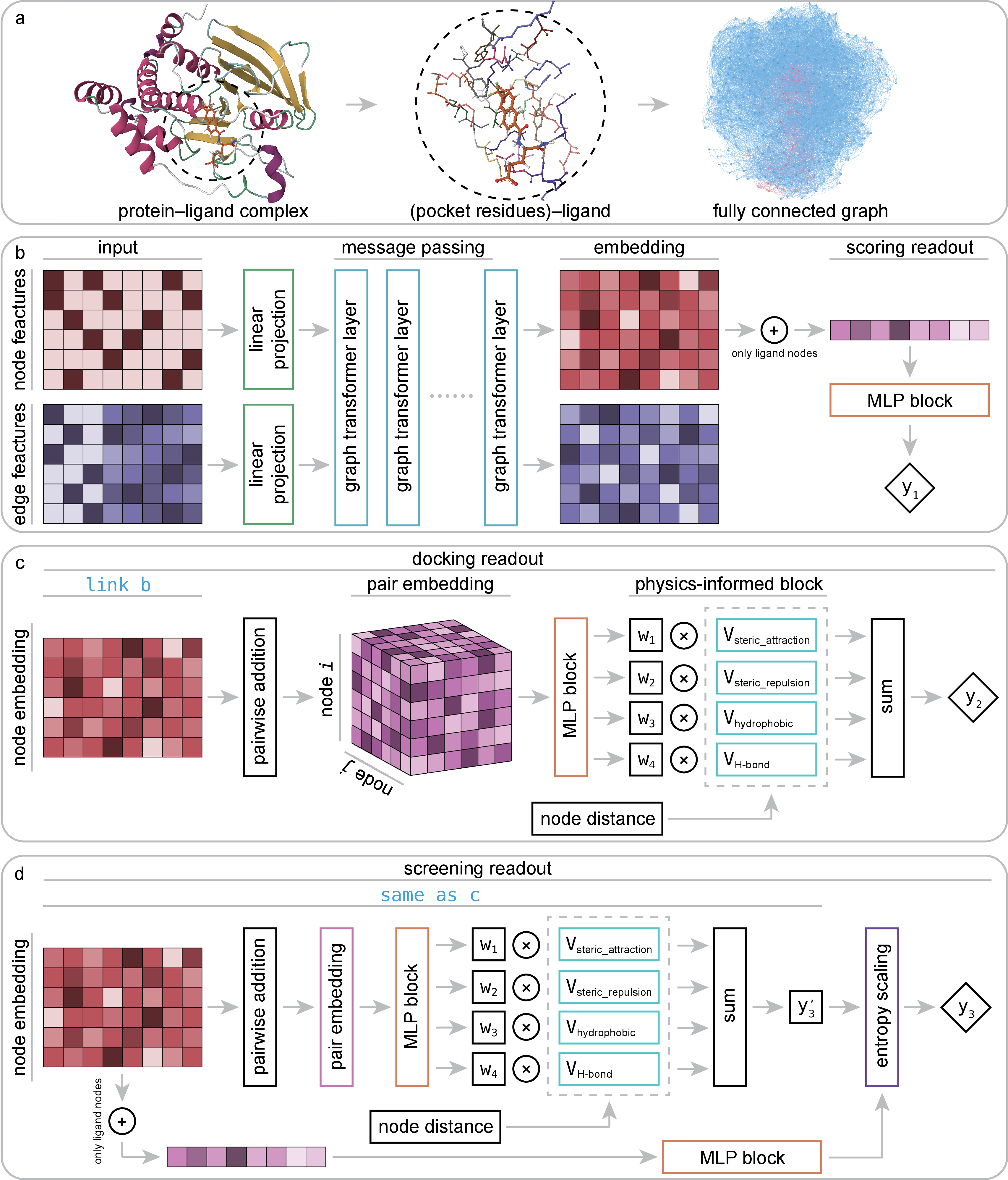}
  \caption{Schematic representation of the multi-objective DeepRLI architecture for protein--ligand interaction prediction. \textbf{a}, The 3D structure around the binding site of the protein--ligand complex is transformed into a fully connected graph with atoms as nodes and interactions as edges, serving as input for the neural network. \textbf{b}, The input graph representation is processed through the AffinityGTN neural network, composed of a linear projection layer, ten graph transformer layers, a ligand-only global pooling layer, and a multi-layer perceptron block containing three fully connected layers, to output a predicted scoring score. \textbf{c}, The node embeddings in AffinityGTN are pairwise added to form pair embeddings, which are then passed through a fully connected layer to yield weights for four physics-informed interaction terms. Finally, all weighted terms are summed to obtain the docking score. \textbf{d}, The screening readout module is similar to the docking readout, except it includes an additional entropy scaling layer, which ultimately outputs the screening score.}
  \label{fgr:deeprli.architecture}
\end{figure}

The complete model architecture of DeepRLI is illustrated in Figure~\ref{fgr:deeprli.architecture}. It accepts a protein--ligand complex with three-dimensional spatial coordinates as input. Note that the receptors we investigate here are proteins, but since the model adopts atoms rather than residues as the fundamental unit, this framework can also easily be extended to other biological macromolecules.

\subsubsection{Molecular Graph Representation}

Generally, binding affinity is associated with the entire system, corresponding to the free energy difference of the system under distinct states. However, if there is no significant change in the protein backbone before and after binding, binding affinity is largely related only to the residues near the pocket. To reduce computational costs, we focus on atoms near the binding site for binding affinity prediction, specifically considering the small molecule and residues within 6.5 Å of it. Instead of only including atoms within a certain cutoff, our approach encompasses entire residues as long as there is a protein--ligand atom pair within 6.5 Å of each other.

In the subsequent step, the structure comprising the selected residues and small molecules is transformed by the model into a graph $\mathcal{G} = (\mathcal{V},\mathcal{E})$ where atoms serve as nodes $\mathcal{V}$ and interactions form edges $\mathcal{E}$ (Figure~\ref{fgr:deeprli.architecture}-a). To preserve the structural information as comprehensively as possible, we assign an edge to every atom pair whose distance is less than 6.5 Å, a reasonable distance cutoff for interatomic interactions. Consequently, such graphs typically consist of hundreds of nodes and tens of thousands of edges. Each node $i$ possesses corresponding atomic features $\alpha_i \in \mathbb{R}^{d_{\text{v}} \times 1}$, and each edge similarly encompasses features $\beta_{ij} \in \mathbb{R}^{d_{\text{e}} \times 1}$ representing the interatomic interaction between nodes $i$ and $j$.

\subsubsection{Graph Transformer with Cosine Envelope}

To achieve adequate expressive power, input node features and edge features are first embedded into a $d$-dimensional hidden space via learnable affine transformations, respectively:
\begin{equation}
  v_i^0 = A \alpha_i + a; \quad e_{ij}^0 = B \beta_{ij} + b,
\end{equation}
where $A \in \mathbb{R}^{d \times d_{\text{v}}}$, $B \in \mathbb{R}^{d \times d_{\text{e}}}$ and $a, b \in \mathbb{R}^d$ are the learnable parameters of the linear projection layers. Our model does not introduce node positional encodings, as atoms in the same context contribute equally to the interaction, and ensuring a unique representation for each node is unnecessary.

\begin{figure}[!hbtp]
  \includegraphics[width=\textwidth]{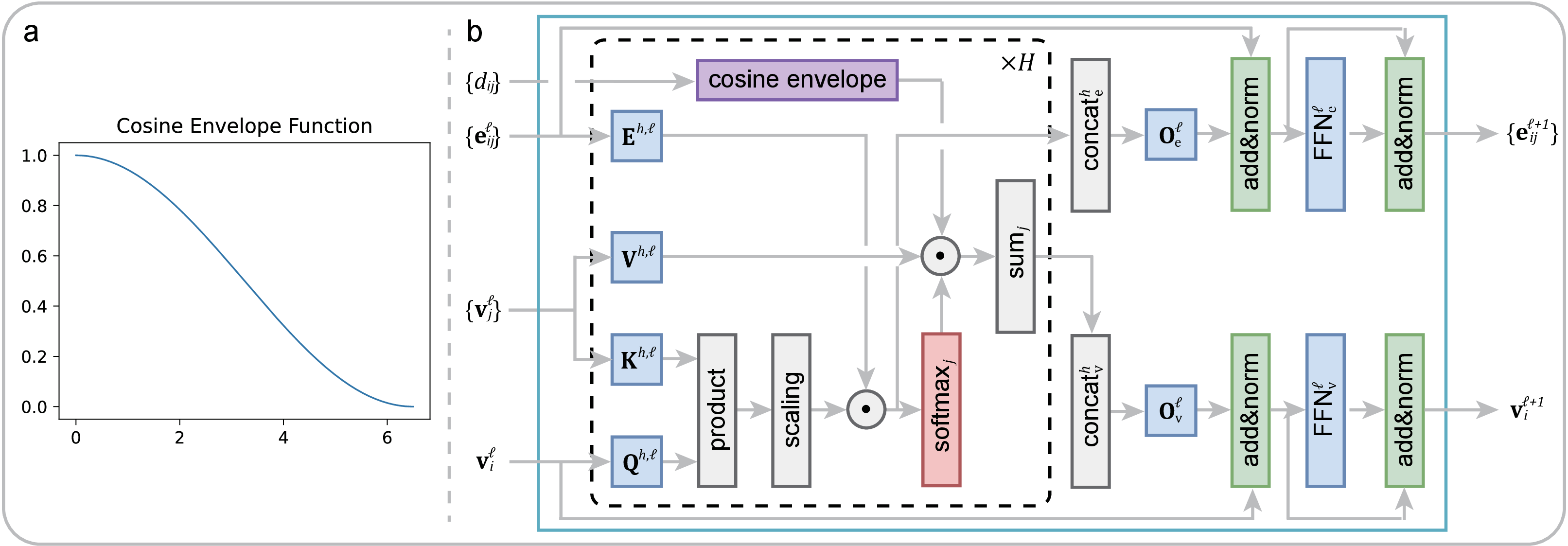}
  \caption{The improved graph transformer used in DeepRLI. \textbf{a}, The curve of cosine envelope function. \textbf{b}, Block diagram of graph transformer with cosine envelope.}
  \label{fgr:graph_transformer_with_consine_envelope}
\end{figure}

Following the initial embedding, hidden node features and edge features undergo updates through ten graph transformer layers (Figure~\ref{fgr:graph_transformer_with_consine_envelope}). Significantly, in our DeepRLI model, we use a refined graph transformer architecture to enhance its applicability to molecular systems. This adaptation is based on the principle that, within a molecular structure, the importance of neighboring atoms to a central atom diminishes with increasing distance, and the contextual representation of the central atom is predominantly influenced by the immediate, proximal atoms. Therefore, we introduce a cosine envelope factor, which is applied to the weights derived from the key-query dot product, modulating them to decay with increasing interatomic distances. The incorporation of this cosine envelope function is crucial, particularly in scenarios of limited training data.  In the absence of this modification, the model may inappropriately focus on learning specific long-distance atomic interactions, potentially leading to overfitting. By implementing this distance-sensitive weighting mechanism, our model more effectively captures the local chemical environment of each atom, thus mitigating the risk of overfitting and enhancing the model's generalizability in drug design applications.

In a single graph transformer layer, the convolution procedure employs the following message-passing scheme: the embedding of a node is updated based on the information from all adjacent nodes and edges, while the embedding of an edge is updated according to the information from its end nodes and itself. It can be expressed as \cite{dwivedi2021generalization}:
\begin{equation}
  \hat{v}_i^{\ell+1} = O_{\text{v}}^{\ell} \Bigg\Vert_{k=1}^H\Big(\sum_{j \in \mathcal{N}_i} c_{i j} w_{i j}^{k, \ell} V^{k, \ell} v_j^{\ell}\Big); \quad \hat{e}_{i j}^{\ell+1} = O_{\text{e}}^{\ell} \Bigg\|_{k=1}^H\Big(\hat{w}_{i j}^{k, \ell}\Big),
\end{equation}
where,
\begin{equation}
  c_{ij} = \frac{1}{2} \cos \left(\frac{\pi d_{ij}}{6.5} + 1\right), \quad w_{i j}^{k, \ell} = \operatorname{softmax}_j\left(\hat{w}_{i j}^{k, \ell}\right), \quad \hat{w}_{i j}^{k, \ell} = \left(\frac{Q^{k, \ell} v_i^{\ell} \cdot K^{k, \ell} v_j^{\ell}}{\sqrt{d_k}}\right) \cdot E^{k, \ell} e_{i j}^{\ell}.
\end{equation}

In the above formulas, $d_{i j}$ is the distance between nodes $i$ and $j$; $\ell$ represents the layer number; $k$ denotes the index of $H$ attention heads; $\|$ signifies concatenation; $\mathcal{N}_i$ refers to neighbor nodes of atom $i$; $Q^{k, \ell}$, $K^{k, \ell}$, $V^{k, \ell} \in \mathbb{R}^{d_k \times d}$ correspond to the query, key, and value generation matrices in the attention mechanism, respectively; $E^{k, \ell} \in \mathbb{R}^{d_k \times d}$ is the linear transformation matrix of edge information, with its projection results used to adjust attention scores; $O_{\text{v}}^{\ell}, O_{\text{e}}^{\ell} \in \mathbb{R}^{d \times d}$ represent the updating functions. The subsequent output $\hat{v}_i^{\ell+1}$, $\hat{e}_{i j}^{\ell+1}$ are each followed by a residual connection and batch normalization layer, a fully-connected layer, and another residual connection and batch normalization layer \cite{dwivedi2021generalization}:
\begin{equation}
  \hat{\hat{v}}_i^{\ell+1} = \operatorname{BatchNorm}(v_i^\ell + \hat{v}_i^{\ell+1}); \quad \hat{\hat{e}}_{ij}^{\ell+1} = \operatorname{BatchNorm}(e_{ij}^\ell + \hat{e}_{ij}^{\ell+1}),
\end{equation}
\begin{equation}
  \hat{\hat{\hat{v}}}_i^{\ell+1} = W_{\text{v},2}^{\ell} \operatorname{ReLU}(W_{\text{v},1}^{\ell} \hat{\hat{v}}_i^{\ell+1}); \quad \hat{\hat{\hat{e}}}_{ij}^{\ell+1} = W_{\text{e},2}^{\ell} \operatorname{ReLU}(W_{\text{e},1}^{\ell} \hat{\hat{e}}_{ij}^{\ell+1}),
\end{equation}
\begin{equation}
  v_i^{\ell + 1} = \operatorname{BatchNorm}(\hat{\hat{v}}_i^{\ell + 1} + \hat{\hat{\hat{v}}}_i^{\ell+1}); \quad e_{ij}^{\ell + 1} = \operatorname{BatchNorm}(\hat{\hat{e}}_{ij}^{\ell + 1} + \hat{\hat{\hat{e}}}_{ij}^{\ell+1}),
\end{equation}
in which $W_{\text{v},1}^{\ell}, W_{\text{e},1}^{\ell} \in \mathbb{R}^{2d \times d}$, $W_{\text{v},2}^{\ell}, W_{\text{e},2}^{\ell} \in \mathbb{R}^{d \times 2d}$, and the BatchNorm operation is
\begin{equation}
  f(x) = \frac{x - \operatorname{E}[x]}{\sqrt{\operatorname{Var}[x] + \epsilon}} \cdot \gamma + \beta,
\end{equation}
where $\operatorname{E}$  signifies expectation for embeddings of nodes or edges, $\operatorname{Var}$  represents the corresponding variance, and $\gamma, \beta \in \mathbb{R}^{d}$ are learnable parameter vectors. The description so far mentioned encapsulates the function of a single graph transformer layer. After iterating through this process ten times, the final node embeddings $v_i^L$ and edge embeddings $e_{ij}^L$ are obtained.

In the subsequent phases, the hidden features $v_i^L$ of the nodes undergo distinct processing through three autonomous downstream networks. This process yields three types of scores: scoring scores, docking scores, and screening scores. The nomenclature of these scores reflects their underlying purposes. Specifically, the scoring scores are tailored for evaluating and ranking crystal structures, the docking scores are optimized for molecular docking processes, and the screening scores are designed for efficiency in virtual screening tasks. This structured approach aims to enhance the precision and applicability of each score to its respective domain within computer-aided drug design.

\subsubsection{Scoring Readout}

The initial intention in the development of the DeepRLI model is to facilitate the prediction of binding affinity, that is, focusing on the accurate quantification of binding free energy values. The corresponding downstream network is the scoring readout. As shown in Figure~\ref{fgr:deeprli.architecture}-b, this part of the neural network is referred to by us as AffinityGTN. The embeddings of the ligand nodes $\mathcal{N}_{\text{lig}}$ obtained after passing through the graph transformer layers are aggregated as the graph's hidden features. This approach is employed because the features associated with affinity are primarily determined by the ligand's environment, and performing global pooling of the entire graph would introduce noise related to residues,
\begin{equation}
x = \sum_{i \in \mathcal{N}_{\text{lig}}} v_i^L.
\end{equation}

Following this, the next step of the model commences, where the hidden graph features $x$ are fed into a multi-layer perceptron (MLP) to generate a scoring score:
\begin{equation}
y_1 = W_3^{\text{r}} \operatorname{ReLU}(W_2^{\text{r}} \operatorname{ReLU}(W_1^{\text{r}} x + b_1^{\text{r}}) + b_2^{\text{r}}) + b_3^{\text{r}},
\end{equation}
where $W_1^{\text{r}} \in \mathbb{R}^{d / 2 \times d}$, $W_2^{\text{r}} \in \mathbb{R}^{d / 4 \times d / 2}$, $W_3^{\text{r}} \in \mathbb{R}^{1 \times d / 4}$ and $b_1^{\text{r}} \in \mathbb{R}^{d / 2}$, $b_2^{\text{r}} \in \mathbb{R}^{d / 4}$, $b_3^{\text{r}} \in \mathbb{R}$ are learnable parameters of linear layers.

\subsubsection{Docking Readout}

It is noteworthy that AffinityGTN is purely data-driven, inferring binding affinity based on the similarity of graph embeddings. This approach, however, introduces a critical deficiency: the model's capability to generalize across a diverse range of molecular structures is inherently constrained by the breadth and variety of the training data. Currently, our knowledge of accurate binding free energy is limited to approximately ten thousand known protein--ligand complex crystal structures. It means that AffinityGTN predominantly learns from data that may not be representative of the entire spectrum of protein--ligand interactions, thereby restricting its understanding to only these biased data and impeding its ability to grasp the more complex, underlying physical principles behind these interactions.

This limitation is particularly relevant in the context of molecular docking and virtual screening tasks. Their objectives often involve estimating the binding scores for structures in weak binding states, which can differ significantly from the configurations of experimentally determined crystal structures. Therefore, enhancing the model's generalization ability to infer on these loose states is of vital importance.

Here, we adopt two approaches together to tackle the challenge of model generalization: data augmentation and the integration of physical constraints. On one hand, data augmentation methodologically broadens the scope of the training set by encompassing a more diverse range of chemical compositions and phase spaces thereof. This expansion ensures a comprehensive coverage of potential scenarios in the model's training phase. On the other hand, more importantly, we incorporate physical constraints into the model. This is achieved by inlaying terms inspired by fundamental physical principles, thereby ensuring that the model's predictions remain consistent with established physical laws.

Drawing inspiration from the methodology employed in PIGNet \cite{moon2022pignet}, our approach in DeepRLI includes the integration of a specialized physical module. The module is specifically designed to account for the interactions between atomic pairs, adding a layer of physical realism to the model's predictive capabilities. The schematic diagram of this approach is demonstrated in Figure~\ref{fgr:deeprli.architecture}-c and d, wherein we detail the workflow of two downstream readout networks. These networks leverage physics-informed blocks to implement a kind of framework that we term ``neural network parameterized potential function''. It effectively strikes a balance between precision in prediction and the capacity for generalization.

The readout module for docking incorporates a physics-informed block that encapsulates four distinct energy terms, as delineated in Figure~\ref{fgr:deeprli.architecture}-c. These terms are extracted from the Vinardo scoring function \cite{quiroga2016vinardo}, an empirical method renowned in the field. They specifically represent four types of interatomic interactions: steric attraction, steric repulsion, hydrophobic interaction, and hydrogen bonding. Notably, the first two terms are integral in accounting for the van der Waals interaction, and their mathematical formulations are presented as follows:
\begin{equation}
  V_{\text{steric\_attraction},ij} = \exp(-(d_{ij}^{\prime} / 0.8)^2),
\end{equation}
\begin{equation}
  V_{\text{steric\_repulsion},ij} = \left\{ \begin{array}{ll}
    d_{ij}^{\prime 2} & \text{if} \ d_{ij}^{\prime} < 0 \\
    0 & \text{if} \ d_{ij}^{\prime} \ge 0 \\
  \end{array} \right..
\end{equation}
In the above formulas, $d_{ij}^{\prime}$ is the reduced distance relative to the atomic surfaces,
\begin{equation}
  d_{ij}^{\prime} = d_{ij} - r_i - r_j,
\end{equation}
where $r$ denotes the van der Waals radius of a atom. Additionally, the remaining two items have similar linear forms:
\begin{equation}
  V_{\text{hydrophobic},ij} = \left\{ \begin{array}{ll}
    1 & \text{if} \ d_{ij}^{\prime} \le 0 \\
    -0.4 (d_{ij}^{\prime} - 2.5) & \text{if} \ 0 < d_{ij}^{\prime} < 2.5 \\
    0 & \text{if} \ d_{ij}^{\prime} \ge 2.5 \\
  \end{array} \right.,
\end{equation}
\begin{equation}
  V_{\text{H-bond},ij} = \left\{ \begin{array}{ll}
    1 & \text{if} \ d_{ij}^{\prime} \le -0.6 \\
    -5 d_{ij}^{\prime} / 3 & \text{if} \ -0.6 < d_{ij}^{\prime} < 0 \\
    0 & \text{if} \ d_{ij}^{\prime} \ge 0 \\
  \end{array} \right.,
\end{equation}
which roughly explain the solvation entropy effect and the dipole-dipole attraction of hydrogen bonds.

We obtain the embedding of any pair of atoms by pairwise adding the node embeddings encoded through the graph transformer, which contains information about the two atoms and their mutual interactions. Subsequently, these pair embeddings are processed through an MLP block, which outputs four weight parameters corresponding to four predefined interaction types. The weighted sum of these four components represents the model's prediction of the interaction between a pair of atoms,
\begin{equation}
  V_{ij} = w_1 V_{\text{steric\_attraction},ij} + w_2 V_{\text{steric\_repulsion},ij} + w_3 V_{\text{hydrophobic},ij} + w_4 V_{\text{H-bond},ij}.
\end{equation}
And the aggregation of the interactions of all atom pairs result in the docking score of the protein--ligand interaction as predicted by the model:
\begin{equation}
  y_2 = \sum_{i<j} V_{ij}.
\end{equation}

\subsubsection{Screening Readout}

In our framework, the screening readout parallels the docking readout in its foundational reliance on a physics-informed block. This similarity notwithstanding, a distinctive feature of the screening readout is the integration of an entropy scaling layer prior to generating the final output. This layer plays a crucial role in compensating for the conformational entropy losses, as delineated in Figure~\ref{fgr:deeprli.architecture}-d. Delving into the specifics, the process involves the transformation of node embeddings through a network analogous to the docking readout, resulting in the derivation of an intermediate variable, denoted as $y_3^{\prime}$. Concurrently, a network akin to the scoring readout is employed to ascertain parameters $w_5$, which is directly applied to scale the number of rotatable bonds $N_{\text{rot}}$ in the ligand. Culminating this process, the screening score, represented as $y_3$, is computed, adhering to the stipulated formula:
\begin{equation}
  y_3 = \frac{y_3^{\prime}}{1 + w_5 N_{\text{rot}}}.
\end{equation}

The above delineates the fundamental architecture of DeepRLI, a deep learning model designed for drug discovery. It takes the three-dimensional structure of a protein--ligand complex as input and, after sophisticated calculations, outputs three scores: a scoring score, a docking score, and a screening score. The essence of these scores is related to the binding free energy $\Delta G_{\text{bind}}$, meaning the smaller the score, the tighter the binding.

\subsection{Input Features}

Node features, $\alpha_i$, and edge features, $\beta_{ij}$, are chemical features extracted from atoms and bonds, respectively, and then transformed into representations suitable for machine learning. Specifically, for the input of our neural network, the node features are represented as 39-dimensional vectors, detailed in Table~\ref{tbl:list_of_node_features}. These vectors include a dimension to differentiate between protein and ligand atoms. The remaining dimensions encapsulate atomic properties derived using the RDKit cheminformatics package \cite{rdkit}. It is important to note that our model's input criteria exclude hydrogen atoms, focusing exclusively on heavy atoms. Consequently, chemical element symbols in our representation do not include hydrogen. Moreover, due to the negligible metal content in the PDBbind dataset \cite{wang2004pdbbind,wang2005pdbbind}, all metal elements are collectively categorized under a single ``Met'' element. Elements not explicitly listed are denoted as ``Unk'' (Unknown). Additionally, the ``degree'' feature in our model quantifies the number of covalent bonds an atom forms with other heavy atoms, effectively representing the number of edges connected to a node. The other attributes of the node features are self-explanatory and contribute to the comprehensive representation of chemical entities in our neural network model.

\begin{table}[!hbt]
  \centering
  \caption{List of node features}
  \label{tbl:list_of_node_features}
  \begin{tabular}{cc}
    \toprule
    Feature Name & Feature Vector \\
    \midrule
    Affiliation & 0 or 1 (for protein or ligand) \\
    Symbol & C, N, O, F, P, S, Cl, Br, I, Met, Unk (one hot) \\
    Hybridization & S, SP, SP2, SP3, SP3D, SP3D2 (one hot) \\
    Formal Charge & -2, -1, 0, 1, 2, 3, 4 (one hot) \\
    Degree & 0, 1, 2, 3, 4, 5 (one hot) \\
    Is a Donor & 0 or 1 \\
    Is an Acceptor & 0 or 1 \\
    Is Negative Ionizable & 0 or 1 \\
    Is Positive Ionizable & 0 or 1 \\
    Is Zn Binder & 0 or 1 \\
    Is Aromatic & 0 or 1 \\
    Is a Hydrophobe & 0 or 1 \\
    Is a Lumped Hydrophobe & 0 or 1 \\
    \bottomrule
  \end{tabular}
\end{table}

The edge features are 39-dimensional vectors, as shown in Table~\ref{tbl:list_of_edge_features}, among which two dimensions are designated for discerning whether the interaction is intermolecular or covalent, and the remainder includes the type of chemical bond and the distance between atoms.  It should be noted that the atomic distance is not encapsulated by a singular dimension, but is instead represented through a series of 33 Gaussian functions, uniformly distributed within a range of 6.5 Å,  each with a width equivalent to the interval. This method of representation results in an expanded distance vector, consisting of multiple values ranging between 0 and 1. Such a multi-valued representation of distance is more effective for the model, facilitating a nuanced utilization of distance data.

\begin{table}[!hbtp]
  \centering
  \caption{List of edge features}
  \label{tbl:list_of_edge_features}
  \begin{tabular}{cc}
    \toprule
    Feature Name & Feature Vector \\
    \midrule
    Is Intermolecular & 0 or 1 \\
    Is Covalent & 0 or 1 \\
    Bond Type & single, double, triple, aromatic (one hot) \\
    Distance & (gaussian smearing, 33) \\
    \bottomrule
  \end{tabular}
\end{table}

\subsection{Datasets}

The training and validation of the DeepRLI model were conducted using datasets that encompass crystal structure-activity data from PDBbind-v2020, supplemented with derived re-docking and cross-docking data. Corresponding to the three training objectives, our dataset also comprises these three parts. The PDBbind database collects a wealth of protein--ligand complex structures and related experimental binding affinity data, making it the most widely used dataset for structure-based protein--ligand binding affinity prediction studies. For the scoring objective, we need to know some structures' precise binding free energy data, which can be directly obtained from PDBbind. However, nearly half of the data in the PDBbind general set are experimental results with only $\text{IC}_{50}$ values or record imprecise $K_{\text{d}}$ (notated as greater than, less than, or approximately equal to). Therefore, we remove these data and retained only the crystal structure data with exact $K_{\text{d}}$ values. And this curated dataset is named PDBbindGS\_HiQ by us.

To enhance the robustness and accuracy of our model in docking and screening tasks, a key requirement is to ensure its adeptness in generalizing to loosely bound structures. To address this, we re-dock the structures from the PDBbind refined set using AutoDock Vina \cite{trott2010autodockvina,eberhardt2021autodock}, thus generating a series of binding conformations. These are compiled into what we have termed the PDBbindRS\_RD dataset, which serves to significantly bolster the model's capability in docking predictions. Given that the exact relative free energy values of these conformations are not precisely known, we posit a correlation between the root-mean-square deviation (RMSD) of these conformations from the original crystallographic structures and their relative free energy. Conformations exhibiting an RMSD of 2 Å or less are hypothesized to possess lower relative free energy, and were thus classified as positive instances (truths). Conversely, those with an RMSD of 4 Å or greater are categorized as negative (decoys). Furthermore, to augment the model's screening proficiency, we initiate a cross-docking protocol involving structures from the PDBbind refined set, thereby creating the PDBbindRS\_CD dataset. This process entails docking various small molecules present in the database with a range of proteins. All conformations resultant from this process are deemed as negative (decoys), providing a comprehensive dataset for enhancing the predictive accuracy of our model in identifying viable drug candidates.

The data unit for training is formed by a collection of structures. Specifically, the minimal input required for training encompasses several components associated with the same protein target: a crystal structure-activity data pair, a randomly selected re-docked positive structure, a randomly selected re-docked negative structure, and a randomly selected cross-docked negative structure, as detailed in Table~\ref{tbl:list_of_data_units}. For a data unit to comply with our criteria, it is essential that it contains at least one instance of these specified data types, corresponding to a particular Protein Data Bank (PDB) identifier. After Intersecting the PDBbindGS\_HiQ, PDBbindRS\_RD, and PDBbindRS\_CD datasets and removing data duplicated with the CASF-2016 benchmark test set, we ultimately obtain 4156 such data units. Additionally, to fully utilize the limited but valuable crystal structure-activity data pairs, we randomly supplement the remaining data from PDBbindGS\_HiQ (7337 items) into the aforementioned data units during training. This approach is implemented to maximize the utility of the available data in our training protocol.

\begin{table}[!hbtp]
  \centering
  \caption{List of data units for training}
  \label{tbl:list_of_data_units}
  \begin{tabulary}{\textwidth}{CCCCC}
    \toprule
    No. & Native ID & Re-docked Positive IDs & Re-docked Negative IDs & Cross-docked Negative IDs \\
    \midrule
    1 & 10gs & [10gs\_27, 10gs\_6] & [10gs\_1, 10gs\_10, 10gs\_11, $\ldots$] & [10gs-1a9q, 10gs-1bn1, 10gs-1bnv, $\ldots$] \\
    2 & 184l & [184l\_1, 184l\_12, 184l\_13, $\ldots$] & [184l\_11, 184l\_14, 184l\_15, $\ldots$] & [184l-1ghy, 184l-1t4v, 184l-3l0v, $\ldots$] \\
    $\vdots$ & $\vdots$ & $\vdots$ & $\vdots$ & $\vdots$ \\
    4156 & 966c & [966c\_1, 966c\_21] & [966c\_11, 966c\_12, 966c\_13, $\ldots$] & [966c-1c1r, 966c-1d09, 966c-1d3d, $\ldots$] \\
    \bottomrule
  \end{tabulary}
  + Supplementary Native IDs (randomly select one for each row): [1afl, 1avn, 1bai, $\ldots$]
\end{table}

\subsection{Training}

\begin{figure}[!hbt]
  \includegraphics[width=\textwidth]{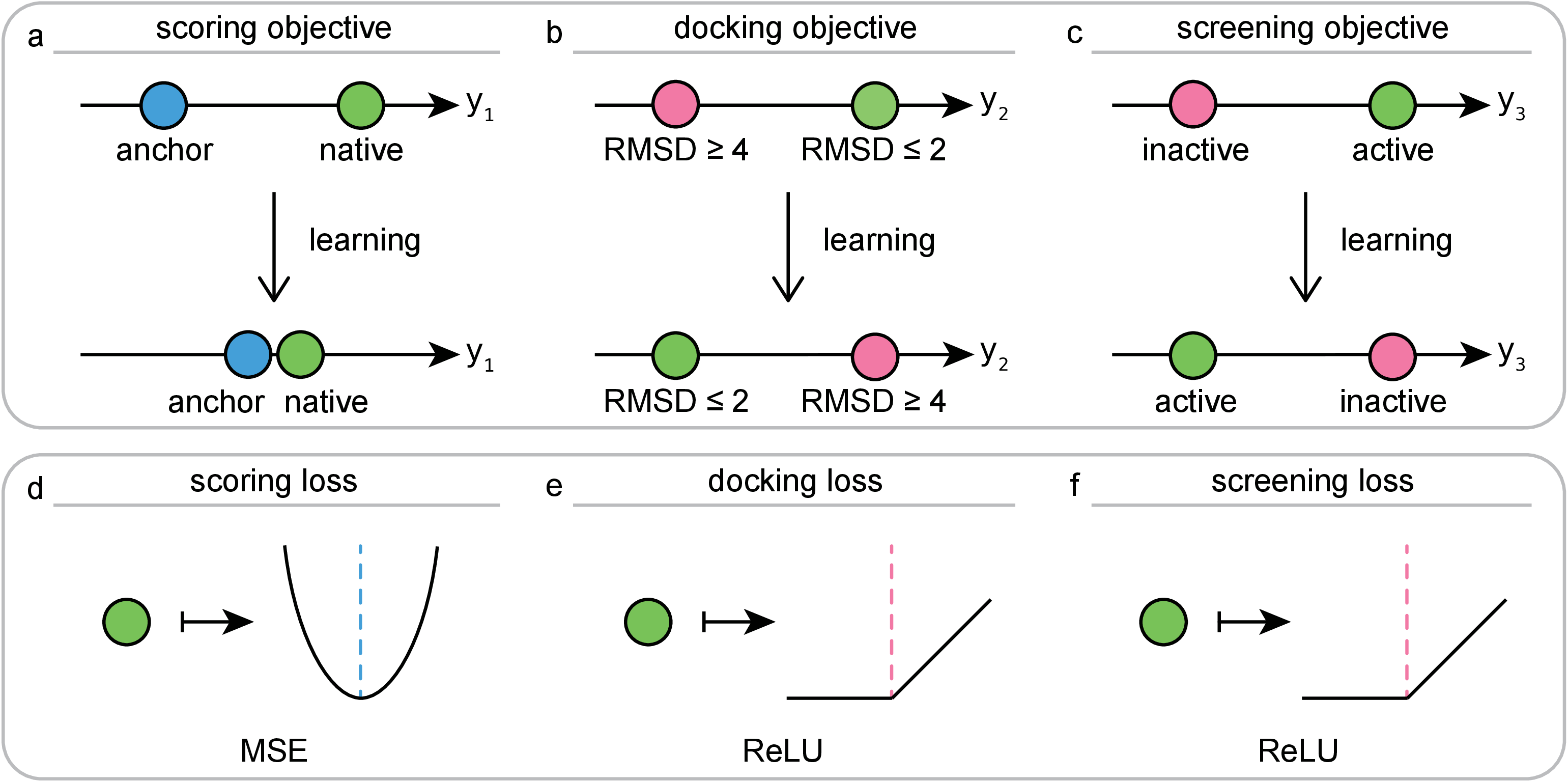}
  \caption{Schematic diagram of the training objectives and corresponding loss functions for the DeepRLI interaction prediction model. \textbf{a}, The training objective for scoring readout is to make the predicted scoring score for native crystal structures close to the experimentally determined binding free energy anchor points. \textbf{b}, The training objective for docking readout is to ensure that the predicted docking score for any pose with $\text{RMSD} \le 2\ \text{\AA}$ from the native crystal structure's small molecule is lower than that for any pose with $\text{RMSD} \ge 4\ \text{\AA}$. \textbf{c}, The training objective for screening readout is to make the predicted screening score for any active ligand lower than that for any inactive decoy. \textbf{d}, Loss function used to achieve the scoring objective. \textbf{e}, Loss function used to achieve the docking objective. \textbf{f}, Loss function used to achieve the screening objective.}
  \label{fgr:deeprli.optimization_objective_and_loss}
\end{figure}

The overarching aim of our model training is the concurrent optimization of predictions for three distinct variables: scoring scores, docking scores, and screening scores. This tripartite goal, depicted in Figure~\ref{fgr:deeprli.optimization_objective_and_loss}, comprises the scoring, docking, and screening objectives. These objectives, while being distinct and relatively independent, are intricately interrelated.

\begin{enumerate}
  \item \textbf{Scoring Objective}: This involves refining the scoring scores to align the model's predictions more closely with the actual relative free energies. Given that the experimental binding free energies (anchors) are available only for the native crystal structures, our focus is on enhancing the accuracy of the model's scoring predictions specifically for these native poses.
  \item \textbf{Docking Objective}: The goal here is to fine-tune the docking scores. The model is trained to yield lower docking scores for poses that closely resemble the native binding pose. Specifically, we aim to achieve lower predicted docking scores for any pose with the RMSD less than 2 Å from the native crystal structure's small molecule compared to poses with the RMSD greater than 4 Å.
  \item \textbf{Screening Objective}: This objective seeks to optimize the screening scores, with a focus on minimizing the scores predicted for active binders. Essentially, the model is calibrated to ensure that the predicted screening scores for any active ligand are lower than those for any inactive decoy.
\end{enumerate}

Through these tailored objectives, our model aims to achieve a nuanced and precise prediction capability, catering to the specific demands of each aspect of the drug design process.

In alignment with the previously delineated objectives, the DeepRLI's loss function comprises three distinct components: scoring loss, docking loss, and screening loss, as articulated in Equations~\ref{eqn:loss_function_1} and \ref{eqn:loss_function_2}:
\begin{IEEEeqnarray}{rCl}
  \mathcal{L} & = & \mathcal{L}_{\text{scoring}} + \mathcal{L}_{\text{docking}} + \mathcal{L}_{\text{screening}} \label{eqn:loss_function_1} \\
  & = & \frac{1}{N} \sum_{i=1}^{N} \left[ \underbrace{\left(y_{\text{native},i}^{\text{pred},1} - y_{\text{native},i}^{\text{true}}\right)^2 + \left(y_{\text{suppl},i}^{\text{pred},1} - y_{\text{suppl},i}^{\text{true}}\right)^2}_{\text{scoring loss}} \right. \nonumber \\
  && \negmedspace {} \left. + \underbrace{\operatorname{max} \left(0, y_{\text{rd-pos},i}^{\text{pred},2} - y_{\text{rd-neg},i}^{\text{pred},2}\right)}_{\text{docking loss}} + \underbrace{\operatorname{max} \left(0, y_{\text{rd-pos},i}^{\text{pred},3} - y_{\text{cd-neg},i}^{\text{pred},3}\right)}_{\text{screening loss}} \right], \label{eqn:loss_function_2}
\end{IEEEeqnarray}
in which ``suppl'', ``rd-pos'', ``rd-neg'', and ``cd-neg'' serve as concise representations for ``supplementary'', ``re-docked positive'', ``re-docked negative'', and ``cross-docked negative'', respectively.

The scoring loss adheres to a conventional methodology, utilizing the Mean Squared Error (MSE) to quantify the discrepancy between the scoring score predicted by the model, denoted as $y^{\text{pred},1}$, and the corresponding experimental binding free energy, $y^{\text{true}}$. 

\begin{figure}[!hbtp]
  \includegraphics[width=0.5\textwidth]{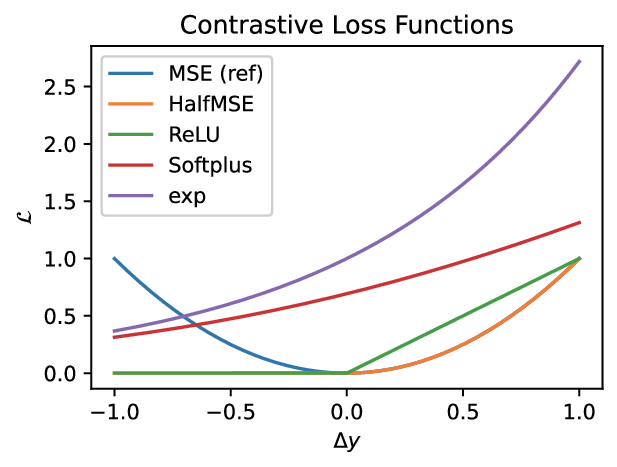}
  \caption{Some contrastive loss functions suitable for docking and screening training objectives of DeepRLI, all of which can optimize the model towards the correct binary relationships, with MSE also drawn in the figure as a reference.}
  \label{fgr:contrastive_loss_functions}
\end{figure}

Conversely, for structures resultant from docking processes, their exact relative free energies remain elusive. However, we can roughly know the relative magnitude of free energy between certain structures. Therefore, we innovatively introduce a contrastive loss function to help achieve docking and screening objectives. The selection of an appropriate contrastive loss function presents multiple viable options, including HalfMSE, ReLU, Softplus, exp, etc., as depicted in Figure~\ref{fgr:contrastive_loss_functions}, with comprehensive derivations available in the ESI. Noteworthy is the characteristic of both HalfMSE and ReLU, which feature a segment on their left spectrum that incurs no loss, thereby ensuring null loss when predictions accurately reflect the true binary relationships. This design effectively circumvents the potential issue of artificially induced gaps in predicted values, a concern prevalent in functions like Softplus and exp. Furthermore, the right extremity of the ReLU function exhibits a more gradual slope compared to HalfMSE, offering a degree of leniency towards certain incorrectly presupposed binary relationships. Consequently, after thorough consideration, ReLU was selected as the most suitable contrastive loss function for our docking and screening objectives, as detailed in Equations~\ref{eqn:loss_function_2}.

In this study, the aforementioned dataset was partitioned into a training set and a validation set in a 9:1 ratio. For optimization, the Adam algorithm was utilized, supplemented with a plateau-based learning rate decay strategy. This approach entails a reduction in the learning rate when no improvement is observed in the validation set loss across a predefined number of consecutive epochs. The training protocol is designed to terminate automatically once the learning rate descends below a specified threshold. The model corresponding to the final epoch was selected as the outcome of the training phase.

\subsection{Evaluation}

To thoroughly assess the efficacy of the versatile DeepRLI model, a comprehensive evaluation was conducted across four critical aspects: scoring, ranking, docking, and screening. The scoring power of a scoring function is defined by its capacity to generate binding scores that linearly correlate with experimental binding data. Similarly, ranking power describes the scoring function's ability to accurately order the known ligands of a specific target protein by their binding affinities, assuming the precise binding poses of these ligands are known. Docking power, on the other hand, refers to the scoring function's proficiency in distinguishing the native ligand binding pose from a set of computer-generated decoys. Finally, screening power is characterized by the scoring function's effectiveness in identifying true binders to a target protein from a collection of random molecules. Each of these capabilities plays a crucial role in the evaluation of the efficacy and reliability of scoring functions.

Considering that the reliance on a single test set is constrained by its specific collection of proteins and small molecules, leading to bias that could potentially skew the model's performance assessment either positively or negatively, the evaluation procedure is diversified to include several widely-recognized benchmark test sets. Among these includes the internal test set of PDBbind, specifically the CASF-2016 benchmark \cite{su2019comparative}---a widely acknowledged standard in this domain. Additionally, external test sets are employed to examine distinct capabilities: the CSAR-NRC HiQ benchmark \cite{dunbar2011csar} for assessing scoring accuracy, the Merck FEP benchmark \cite{schindler2020largescale} for evaluating ranking efficacy, and the LIT-PCBA benchmark \cite{trannguyen2020litpcba} for screening proficiency. This multi-faceted approach ensures a more balanced and thorough evaluation of the DeepRLI model's performance across various scenarios.

\textbf{CASF-2016 benchmark.} The Comparative Assessment of Scoring Functions (CASF) benchmark was created by Cheng \textit{et al.} and first published in 2009 as CASF-2007 \cite{cheng2009comparative}. It has since been maintained and updated, with subsequent releases of CASF-2013 \cite{li2014comparative1,li2014comparative2} and the latest CASF-2016 version. In a nutshell, this dataset consists of different protein--ligand complexes with high-resolution crystal structures and reliable binding constants, obtained through systematically non-redundant sampling from the PDBbind database. And it is used to evaluate the performance of scoring functions regarding protein--ligand binding in the four previously mentioned aspects. Specifically, the CASF-2016 benchmark, which is the focus of this study, comprises an array of 285 high-quality protein--ligand complex crystal structures accompanied by reliable binding constants. Notably, these structures cover 57 different targets, each with five structures bound to different ligands. The structure-activity data pairs can be used for scoring and ranking capability assessment. Moreover, to meet the requirements for docking and screening ability tests, CASF-2016 also includes protein--ligand binding poses (decoys) generated by molecular docking programs. For each protein--ligand complex, a decoy set containing up to 100 ligand binding poses is generated for docking ability assessment. Additionally, each protein is cross-docked with another 280 ligands to obtain a larger decoy set suitable for screening ability assessment.

\textbf{CSAR-NRC HiQ benchmark.} The Community Structure-Activity Resource (CSAR)-National Research Council of Canada (NRC) High-Quality (HiQ) benchmark was proposed in 2010 by Dunbar and many other researchers. CSAR aims to collect data from industry and academia that can be used to improve docking and scoring computational methods. The CSAR-NRC HiQ benchmark primarily serves to evaluate the efficacy of various scoring algorithms. Originally, the dataset encompassed 343 distinct protein--ligand complexes, divided into two sets: set1 with 176 and set2 with 167 complexes. Subsequently, in 2011, the dataset was expanded to include an additional set, set3, comprising 123 complexes, thereby augmenting the total count to 466 structure-activity datasets. A critical aspect to consider is the substantial overlap of complex structures within the CSAR-NRC HiQ dataset and our training set, PDBbindGS\_HiQ. To mitigate the risk of artificially inflating the scoring performance of our model due to this redundancy, we have elected to exclude these overlapping complexes. This adjustment results in a revised dataset composition, with set1, set2, and set3 now containing 50, 36, and 75 complexes, respectively. Furthermore, this benchmark is often not evaluated in its complete state in other published works, for example, parts overlapping with the entire PDBbind general set are removed \cite{shen2023ageneralized}. For ease of comparison, we also evaluate it on the same datasets as those works.

\textbf{Merck FEP benchmark.} Accurately ranking small molecules with subtle differences in binding efficacy to specific proteins plays an important role in the hit-to-lead and lead optimization stages of drug discovery. To address this challenging task, rigorous free energy simulation methods such as free energy perturbation (FEP), thermodynamic integration (TI), and $\lambda$-dynamics are employed for this purpose \cite{chipot2007free}, among which Schr{\"o}dinger FEP+ \cite{wang2015accurate} is currently recognized as a mature and reliable program. In 2020, Merck KGaA published a benchmark for the assessment of relative free energy calculations, namely the Merck FEP benchmark, and tested the effects of large-scale prospective applications of FEP+ \cite{schindler2020largescale}. This dataset collected a total of 8 pharmaceutical targets and 264 active ligands, with ligands belonging to a specific target having analogous skeletons and nuanced structural variations. Therefore, it can be used to further evaluate the ranking power of our model, especially the possibility of its application in hit-to-lead and lead optimization.

\textbf{LIT-PCBA benchmark.} The LIT-PCBA is a recent benchmark specifically designed for the comparative evaluation of virtual screening. Compared to past analogous test sets, it relies on experimental biological activity data from the PubChem BioAssay database to support its crafting, thereby minimizing the presence of false positive and false negative compound data, which is common in past benchmarks due to insufficient experimental data and random selection of decoys. Moreover, it maintains a similar range and distribution of molecular properties for both active and inactive compounds, avoiding inappropriate evaluations of virtual screening methods due to inherent chemical biases. Therefore, the LIT-PCBA benchmark is currently a suitable dataset for evaluating the screening power of machine learning-based scoring functions. Notably, it includes 15 targets, 7955 active compounds, and 2644022 inactive compounds. Such a hit rate (the ratio of active to inactive compounds) accurately reflects the real-world virtual screening scenario, greatly aiding in our further understanding of DeepRLI's screening effectiveness in challenging tasks. Since our model is based on 3D complex structures and currently lacks the capability for conformational sampling, in our tests, we first employ Glide SP \cite{friesner2004glide,halgren2004glide} to ascertain various binding poses of small molecules with proteins, and then screen molecules guided by the predictions of DeepRLI.

In this work, several DeepRLI models are trained, demonstrating comparable performance across subsequent evaluations. Although an ensemble of multiple models could bring a slight performance improvement, considering computational efficiency, we opt to use a single model in the production phase to ensure precise and efficient inference. Therefore, the performance evaluation phase focuses on the test results of an individual DeepRLI model. To benchmark our method's inference capabilities, we conduct a comparative analysis with existing scoring functions. This comparison particularly targeted those functions with comprehensive, detailed evaluation results available, such as the array of scoring functions detailed in CASF-2016 and the variety of scoring models discussed in the GenScore publication. Note that the variability in GenScore's performance across different hyperparameter settings, and the GT\_1.0 model is selected as our baseline for comparison. Further enriching our comparative investigation, we include results from the PIGNet model, another machine learning method inspired by physics-based principles, as well as the PLANET model \cite{zhang2023planet}, known for its ability to expedite virtual screening processes without necessitating binding poses. Baseline data for each model is directly sourced from relevant literature. And the models lacking benchmark-specific data are excluded from certain comparative analysis.

\section{Results}

\subsection{Performance}

In this subsection, the performance of the DeepRLI model is systematically evaluated across the previously mentioned benchmarks. 

\subsubsection{Evaluation on CASF-2016 Benchmark}

\begin{figure}[!hbtp]
  \includegraphics[width=\textwidth]{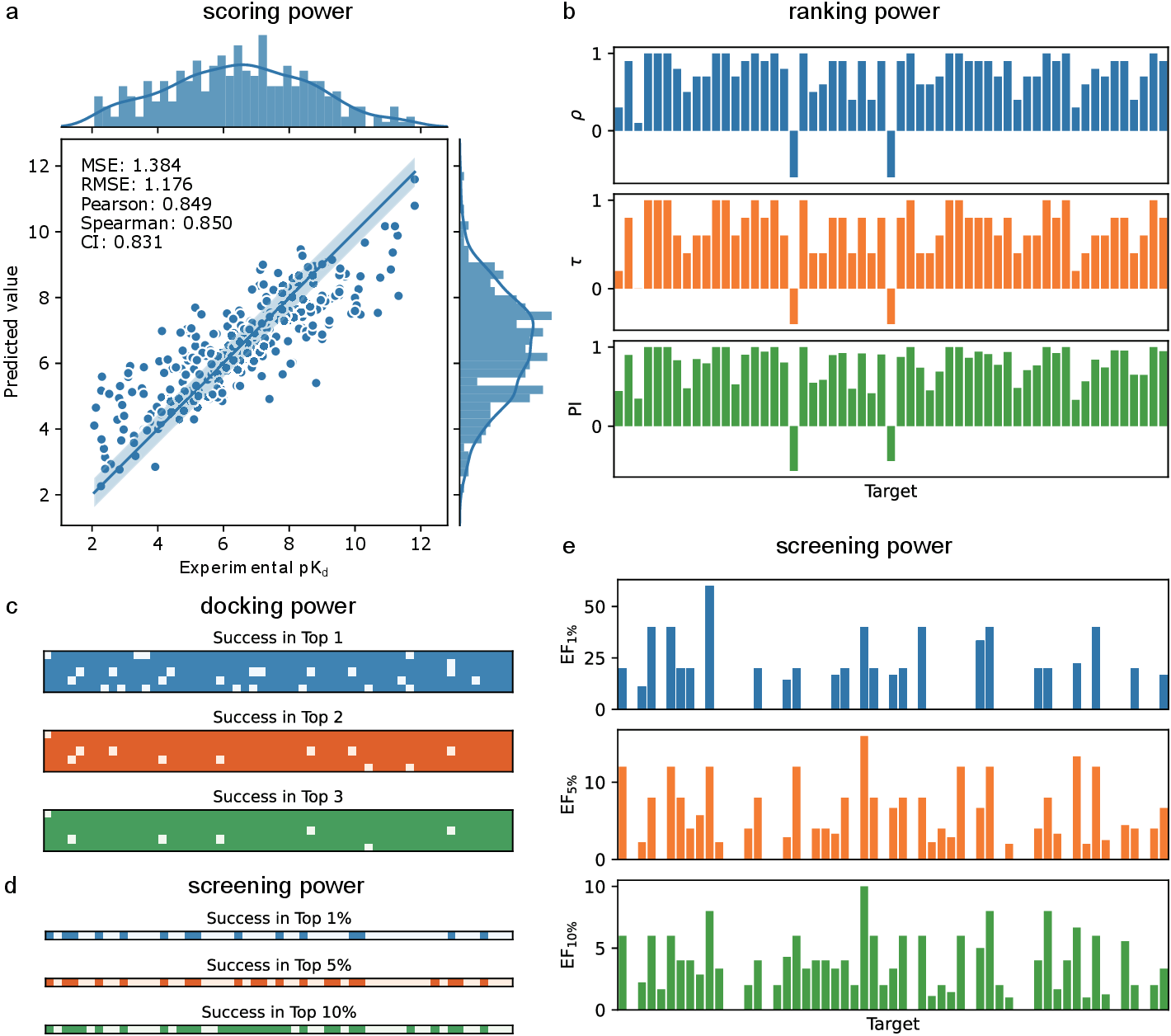}
  \caption{Performance of DeepRLI in scoring, ranking, docking, and screening on the CASF-2016 benchmark. \textbf{a}, A correlation scatter plot depicting DeepRLI prediction of the experimental $pK_{\text{d}}$ values for protein--ligand complexes. The light blue area surrounding the diagonal line represents the range of thermal fluctuations, specifically $\pm0.434$, and data points falling within this range can be considered to be in good agreement with the experimental data. \textbf{b}, Bar plots for three ranking metrics demonstrate DeepRLI's ability to rank active small molecules of various targets. The targets are arranged from left to right in alphabetical order of their PDB IDs. \textbf{c}, Three heatmaps composed of 285 ($5 \times 57$) small squares, where the squares from left to right and top to bottom correspond to complexes arranged in alphabetical order of their PDB IDs. Colored squares represent successfully docked complexes, i.e., those where one of the top $n$ (1, 2, 3) poses predicted by DeepRLI have an RMSD less than 2; in contrast, uncolored squares represent failures. \textbf{d}, Similar to \textbf{c}, these heatmaps show whether active ligands are present in the top $\alpha$ (1\%, 5\%, 10\%) small molecules predicted by DeepRLI.  \textbf{e}, Similar to \textbf{b}, the three bar graphs demonstrate DeepRLI's ability to enrich active small molecules in the top $\alpha$ (1\%, 5\%, 10\%) across various targets.}
  \label{fgr:casf-2016_performance}
\end{figure}

\begin{figure}[!hbtp]
  \includegraphics[width=0.95\textwidth]{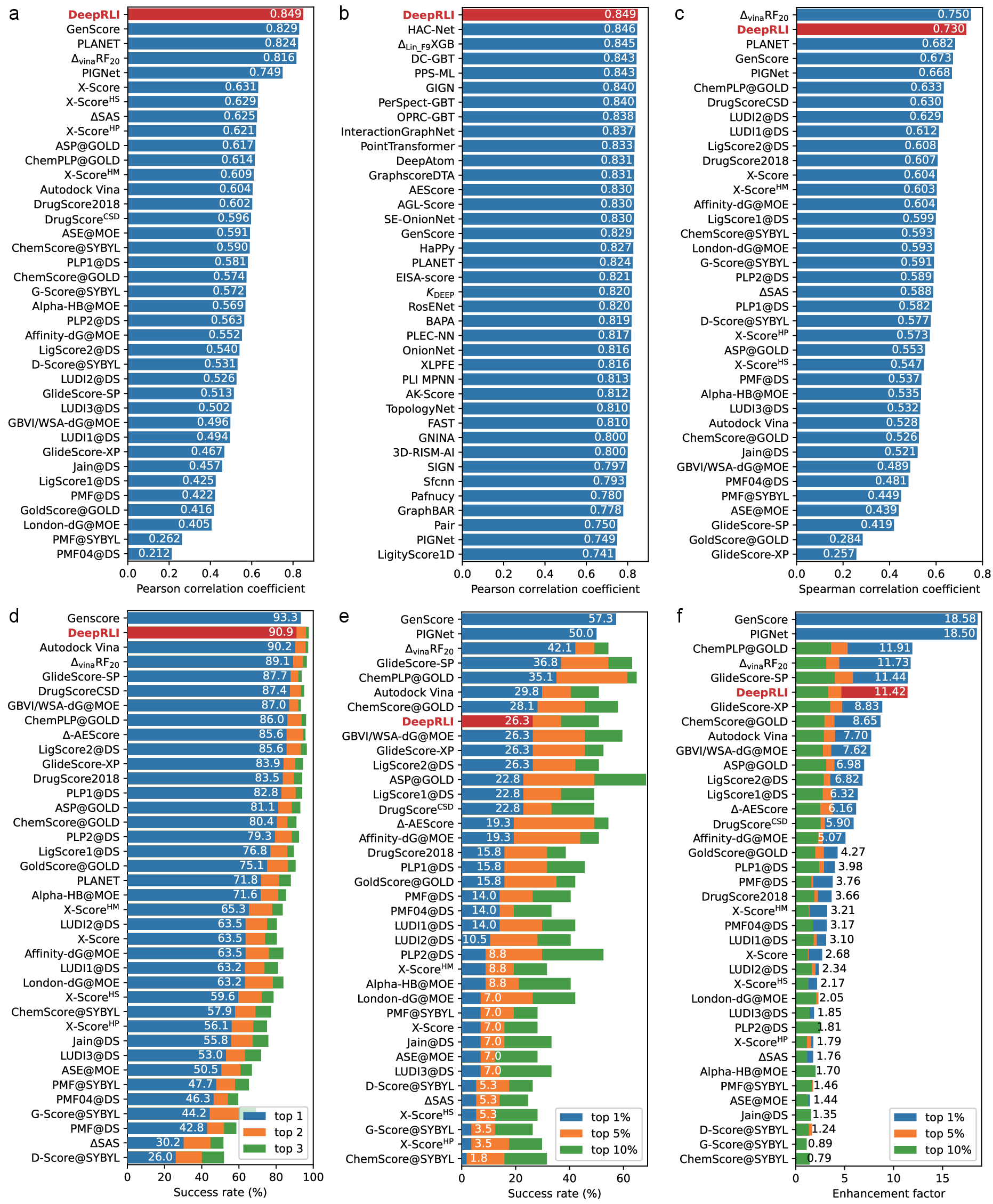}
  \caption{A series of leaderboards comparing various performance metrics of many scoring functions. \textbf{a}, A leaderboard ranked by Pearson correlation coefficient, indicating scoring power. \textbf{b}, Similar to \textbf{a}, but compared with some representative machine learning-based scoring functions. \textbf{c}, A leaderboard ranked by Spearman correlation coefficient, manifesting ranking power. \textbf{d}, A leaderboard ranked by success rate calculated at the top 1 level, demonstrating docking power. \textbf{e}, A leaderboard ranked by success rate calculated at the top 1\% level, demonstrating forward screening power. \textbf{f}, A leaderboard ranked by enhancement factor calculated at the top 1\% level, also reflecting forward screening power.}
  \label{fgr:casf-2016_leaderboards}
\end{figure}

In our initial evaluation, we assess the efficacy of DeepRLI using the CASF-2016 benchmark, which is a comprehensive and widely recognized standard in the field. This benchmark encompasses three distinct structural categories: crystal structures, re-docked structures, and cross-docked structures. Crystal structures are pivotal for gauging the scoring and ranking capabilities of the algorithm. In contrast, re-docking and cross-docking structures play a crucial role in examining the algorithm's proficiency in docking and screening processes, respectively. All pertinent results from this assessment are systematically detailed and displayed in Figure~\ref{fgr:casf-2016_performance}, Figure~\ref{fgr:casf-2016_leaderboards}, Table~\ref{tbl:performance_on_casf-2016.scoring_ranking_docking_power}, and Table~\ref{tbl:performance_on_casf-2016.screening_power}.

\noindent \textbf{Scoring Power}

The scoring power of a model refers to its prediction accuracy of the binding free energy. This is typically assessed by examining the correlation between the computational scores generated by the model and the corresponding experimental data. To quantify this relationship, several statistical metrics are commonly employed. These include the Mean Square Error (MSE) and Root Mean Square Error (RMSE), which measure the average magnitude of the errors in predictions. Additionally, the Pearson correlation coefficient ($R_{\text{p}}$) and Spearman correlation coefficient ($\rho$) are used to assess the linear and rank-order correlations, respectively, between predicted scores and experimental outcomes. The Concordance Index (CI) is another metric offering a measure of the ranking correctness \cite{ozturk2018deepdta}.

Our DeepRLI model shows a strong correlation between the predicted binding affinities for 285 crystal structures in the CASF-2016 dataset and the experimental $pK_{\text{d}}$ data (Figure~\ref{fgr:casf-2016_performance}-a), with an MSE of 1.384, an RMSE of 1.176, $R_{\text{p}}$ of 0.849, $\rho$ of 0.850 and CI of 0.831. In Figures~\ref{fgr:casf-2016_leaderboards}-a and \ref{fgr:casf-2016_leaderboards}-b, we compare the scoring performance of DeepRLI with other scoring functions. Figure~\ref{fgr:casf-2016_leaderboards}-a mainly includes scoring functions from CASF-2016, most of which are traditional methods; Figure~\ref{fgr:casf-2016_leaderboards}-b consists entirely of machine learning-based methods developed in recent years \cite{kyro2023hacnet,yang2022delta,liu2022dowker,liu2023persistent,yang2023geometric,zhenyu2021persistent,wee2021ollivier,jiang2021interactiongraphnet,wang2021apoint,li2019deepatom,wang2023graphscoredta,meli2021learning,nguyen2019aglscore,wang2021seonionnet,shen2023ageneralized,zhang2023amultiperspective,zhang2023planet,rana2022eisascore,jimenez2018kdeep,hassan-harrirou2020rosenet,seo2021binding,Wojcikowski2018development,zheng2019onionnet,dong2022xlpfe,volkov2022on,kwon2020akscore,cang2017topologynet,jones2021improved,francoeur2020three,osaki20223drismai,li2021structureaware,wang2022sfcnn,stepniewska-dziubinska2018development,son2021development,zhu2020binding,moon2022pignet}, most of which are structure-based \cite{meli2022scoring}. In these scoring function rankings, DeepRLI is the highest, reaching the current state-of-the-art level.

\noindent \textbf{Ranking Power}

In the evaluation of scoring capabilities, we incorporated the analysis of several metrics pertinent to ranking. These metrics were calculated across the whole crystal structure test set. Notably, in the context of the CASF assessment, ``ranking power'' is specifically defined as the proficiency in ordering known active ligands against a particular biological target. Nevertheless, it still indicates a positive correlation between scoring and ranking abilities; typically, a robust scoring ability is indicative of a similarly robust ranking ability. To quantitatively measure the ranking power, three primary metrics are utilized: the Spearman correlation coefficient ($\rho$), the Kendall correlation coefficient ($\tau$), and the Predictive Index (PI) \cite{su2019comparative}.

In Figure~\ref{fgr:casf-2016_performance}-b, the ranking efficacy of DeepRLI is demonstrated through its performance in ranking five active small molecules across each of the 57 targets within the CASF-2016 dataset. Notably, the model achieved a perfect prediction score (with all indicators at 1) for several targets, indicating an exact match between the predicted and actual ordering of molecules. For the majority of the targets, the model's predictions exhibited a positive correlation with the actual rankings, as evidenced by scores exceeding 0.5. However, challenges arose in the case of two specific targets, identified by PDB IDs 2ZCQ and 3G0W, where the model's predictions were inversely correlated with the actual data. Further analysis revealed that these discrepancies could be attributed to the presence of multiple ligands with closely similar $pK_{\text{d}}$ values, complicating the task of accurate ranking. Upon aggregating the results across all 57 targets, the overall ranking capability of DeepRLI was quantified, with $\rho$ of 0.730, $\tau$ of 0.660, and PI of 0.757. As delineated in the leaderboard in Figure~\ref{fgr:casf-2016_leaderboards}-c, DeepRLI's ranking performance is positioned at the forefront, aligning with the current state-of-the-art in the field.

\begin{table}[!hbtp]
  \caption{The scoring, ranking and docking powers of several representative scoring functions on the CASF-2016 benchmark. The data for the first 5 methods are from Su \textit{et al.} \cite{su2019comparative}, while for other methods except DeepRLI, the data are from their respective original literatures. The best results in each column are highlighted in bold}
  \label{tbl:performance_on_casf-2016.scoring_ranking_docking_power}
  \begin{tabular}{l cc ccc ccc}
    \toprule
    \multirow{2}{*}{Method} & \multicolumn{2}{c}{Scoring Power} & \multicolumn{3}{c}{Ranking Power} & \multicolumn{3}{c}{Docking Power} \\
    \cmidrule(lr){2-3} \cmidrule(lr){4-6} \cmidrule(lr){7-9}
    & RMSE & $R_{\text{p}}$ & $\rho$ & $\tau$ & PI & $\text{SR}_{1} (\%)$ & $\text{SR}_{2} (\%)$ & $\text{SR}_{3} (\%)$ \\
    \midrule
    Vina \cite{trott2010autodockvina,eberhardt2021autodock} & 1.73 & 0.604 & 0.528 & 0.453 & 0.557 & 90.2 & 95.8 & 97.2 \\
    Glide SP \cite{friesner2004glide,halgren2004glide} & 1.89 & 0.513 & 0.419 & 0.374 & 0.425 & 87.7 & 91.9 & 93.7 \\
    Glide XP \cite{friesner2006extra} & 1.95 & 0.467 & 0.257 & 0.227 & 0.255 & 83.9 & 90.2 & 94.4 \\
    X-Score \cite{wang2002further} & 1.69 & 0.631 & 0.604 & 0.529 & 0.638 & 63.5 & 74.0 & 80.4 \\
    $\Delta_{\text{vina}}\text{RF}_{20}$ \cite{wang2017improving} & 1.26 & 0.816 & \textbf{0.750} & \textbf{0.686} & \textbf{0.761} & 89.1 & 94.4 & 96.5 \\
    $\Delta_{\text{Lin\_F9}}$XGB \cite{yang2022delta} & 1.24 & 0.845 & 0.704 & 0.625 & -- & 86.7 & -- & -- \\
    AEScore \cite{meli2021learning} & 1.22 & 0.830 & 0.640 & 0.550 & 0.670 & 35.8 & 54.4 & 60.4 \\
    $\Delta$-AEScore \cite{meli2021learning} & 1.34 & 0.790 & 0.590 & 0.520 & 0.610 & 85.6 & 94.4 & 95.8 \\
    PLANET \cite{zhang2023planet} & 1.25 & 0.824 & 0.682 & -- & -- & 71.8 & 81.6 & 87.9 \\
    PIGNet \cite{moon2022pignet} & -- & 0.749 & 0.668 & -- & -- & -- & -- & -- \\
    GenScore \cite{shen2023ageneralized} & -- & 0.829 & 0.673 & -- & -- & \textbf{93.3} & -- & -- \\
    DeepRLI & \textbf{1.18} & \textbf{0.849} & 0.730 & 0.660 & 0.757 & 90.9 & \textbf{96.1} & \textbf{97.5} \\
    \bottomrule
  \end{tabular}
\end{table}

\begin{table}[!hbtp]
  \caption{The screening power of several representative scoring functions on the CASF-2016 benchmark. The data for the first 5 methods are from Su \textit{et al.} \cite{su2019comparative}, while for other methods except DeepRLI, the data are from their respective original literatures. The best results in each column are highlighted in bold}
  \label{tbl:performance_on_casf-2016.screening_power}
  \begin{tabular}{lcccccc}
    \toprule
    \multirow{2}{*}{Model} & \multicolumn{6}{c}{Screening Power} \\
    \cmidrule(lr){2-7}
    & $\text{SR}_{1\%} (\%)$ & $\text{SR}_{5\%} (\%)$ & $\text{SR}_{10\%} (\%)$ & $\text{EF}_{1\%}$ & $\text{EF}_{5\%}$ & $\text{EF}_{10\%}$ \\
    \midrule
    Vina & 29.8 & 40.4 & 50.9 & 7.70 & 4.01 & 2.87 \\
    Glide SP & 36.8 & \textbf{54.4} & \textbf{63.2} & 11.44 & \textbf{5.83} & \textbf{3.98} \\
    Glide XP & 26.3 & 45.6 & 52.6 & 8.83 & 4.75 & 3.51 \\
    X-Score & 7.0 & 15.8 & 28.1 & 2.68 & 1.31 & 1.23 \\
    $\Delta_{\text{vina}}\text{RF}_{20}$ & 42.1 & 49.1 & 54.4 & 11.73 & 4.43 & 3.10 \\
    $\Delta_{\text{Lin\_F9}}$XGB & 40.4 & -- & -- & 12.6 & -- & -- \\
    AEScore & -- & -- & -- & -- & -- & -- \\
    $\Delta$-AEScore & 19.3 & 49.1 & 54.4 & 6.16 & 3.76 & 2.48 \\
    PLANET & -- & -- & -- & -- & -- & -- \\
    PIGNet & 50.0 & -- & -- & 18.5 & -- & -- \\
    GenScore & \textbf{57.3} & -- & -- & \textbf{18.58} & -- & -- \\
    DeepRLI & 26.3 & 36.8 & 50.9 & 11.42 & 4.65 & 3.30 \\
    \bottomrule
  \end{tabular}
\end{table}

\noindent \textbf{Docking Power}

The concept of docking power pertains to the proficiency of a scoring function in accurately identifying the native binding pose within a diverse array of protein--ligand conformational states. In CASF-2016, each of the 285 complexes has nearly 100 decoy conformations sampled by various docking programs. And scoring functions evaluate and rank the 285 groups of conformations individually. Notably, a scoring function demonstrating optimal docking ability tends to assign higher ranks to those conformations that closely resemble the binding pose of the complex's native crystal structure. Therefore, the quantitative metric for measuring docking ability is the success rate of having conformations within the top $n$ (1, 2, 3) ranks whose RMSD from the native ligand pose is less than 2 Å.

\begin{figure}[!hbtp]
  \includegraphics[width=0.6\textwidth]{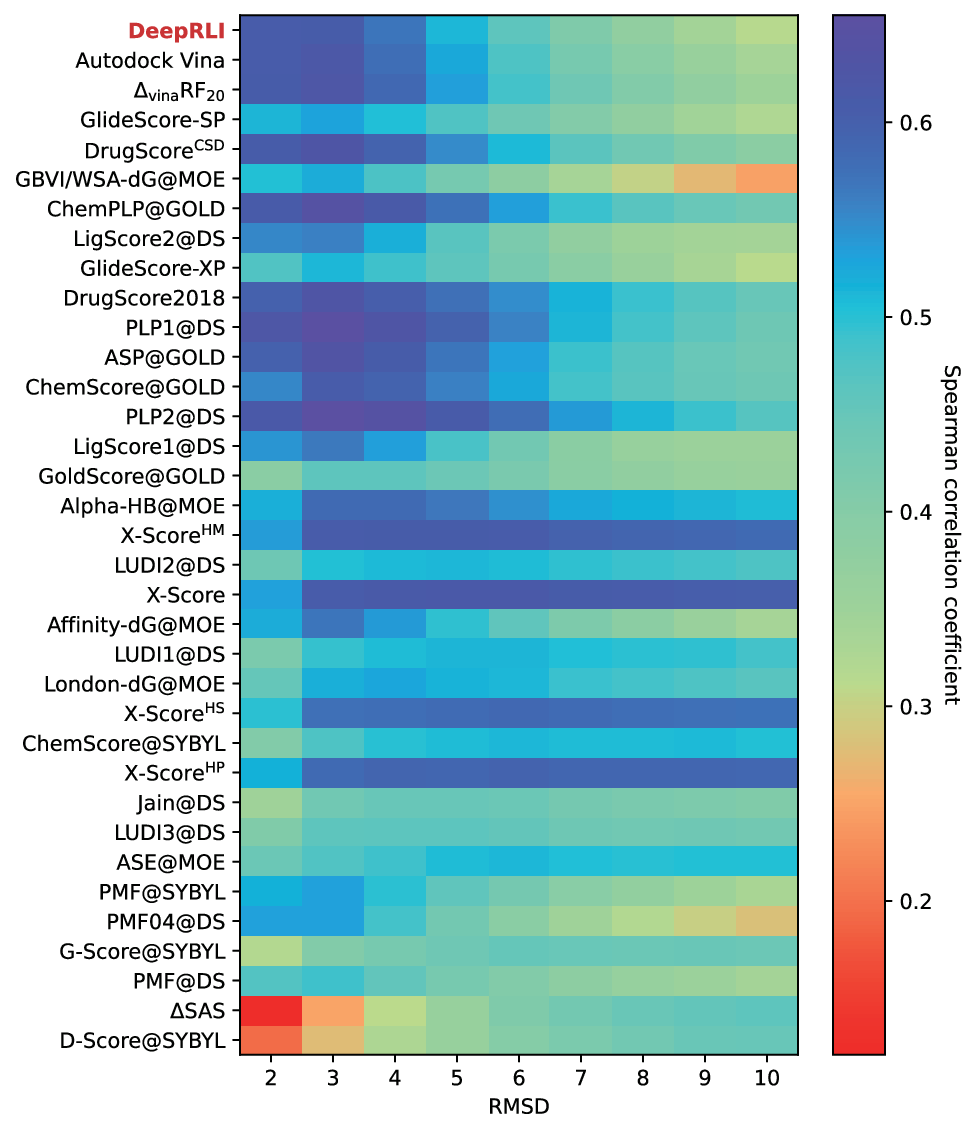}
  \caption{A heatmap displaying the binding funnel landscapes of scoring functions. The ticks on the x-axis refer to the ranges of RMSDs (for example, $0-2$ Å, $0-3$ Å, etc.), and the corresponding blocks indicate the Spearman correlation coefficient between the RMSD values and the binding scores calculated by scoring functions for all ligand poses within these ranges.}
  \label{fgr:casf-2016_docking_funnel_analysis}
\end{figure}

We evaluate the docking performance of DeepRLI on a dataset comprising 285 protein--ligand systems from CASF-2016, as depicted in Figure~\ref{fgr:casf-2016_performance}-c. The results, predominantly represented by dark areas on the heatmap, suggest a high success rate in docking most of the complexes. Specifically, the top 1, 2, and 3 docking success rates achieved by our model are 90.9\%, 96.1\%, and 97.5\%, respectively. Notably, achieving a top 1 success rate exceeding 90\% is a remarkable outcome, positioning our method among the leading approaches in terms of docking capabilities, as demonstrated in Figure~\ref{fgr:casf-2016_leaderboards}-d.

Furthermore, we conduct a binding funnel analysis for DeepRLI, presented in Figure~\ref{fgr:casf-2016_docking_funnel_analysis}. This analysis reveals a strong correlation between the docking scores predicted by DeepRLI and the RMSD values, particularly within a shorter RMSD range (e.g., RMSD < 5 Å). This correlation manifests in the formation of a funnel landscape, indicative of not only the model's high docking accuracy but also its efficiency in docking procedures.

\noindent \textbf{Screening Power}

Screening power denotes the efficacy of a scoring function in accurately identifying potential ligands that exhibit strong binding affinity to a specific protein within a diverse pool of small molecules. CASF-2016 obtained the structures of each of the 57 proteins bound to 280 other small molecules through cross-docking \cite{su2019comparative}. It is important to note that cross-binders do exist, meaning that certain proteins may have more than five true binders, and the goal of screening is to enrich all of these binders. The screening power is quantitatively measured by the success rate in pinpointing the highest-affinity binders within the top 1\%, 5\%, or 10\% of the ranked small molecules. Additionally, the enhancement factors at these top percentile levels also serve as critical metrics for evaluation.

The evaluation of DeepRLI's screening efficacy across 57 proteins within the CASF-2016 framework is depicted in Figures~\ref{fgr:casf-2016_performance}-d and \ref{fgr:casf-2016_performance}-e. While the overall performance of the screening process is moderate, the model exhibit notable proficiency in enriching the majority, even all, active ligands at the forefront for specific targets, notably those with PDB IDs 2P15 and 3EJR. In terms of quantifiable metrics, the top 1\%, 5\%, and 10\% screening success rates of our model are 26.3\%, 36.8\%, and 50.9\%, respectively; and the corresponding enhancement factors are 11.42, 4.65, and 3.30, separately. The screening capability rankings, as illustrated in Figures~\ref{fgr:casf-2016_leaderboards}-e and \ref{fgr:casf-2016_leaderboards}-f, indicate that DeepRLI's performance ranks competitively among traditional scoring functions. However, it does not yet match the efficacy of the leading-edge machine learning-based methodologies. Notably, further assessments conducted on other virtual screening test sets have demonstrated that DeepRLI's screening performance aligns with the forefront of contemporary machine learning-based approaches. This finding underscores DeepRLI's robust generalization capabilities in virtual screening.

To comprehensively demonstrate the performance level of DeepRLI, we have listed in Table~\ref{tbl:performance_on_casf-2016.scoring_ranking_docking_power} and \ref{tbl:performance_on_casf-2016.screening_power} the scoring, ranking, docking, and screening powers of some representative scoring functions on CASF-2016. As can be seen, DeepRLI exhibits robust overall performance. Notably, its screening capability aligns with renowned traditional scoring functions such as Vina, Glide SP, and Glide XP. However, DeepRLI excels in other domains, demonstrating cutting-edge proficiency, particularly in scoring and ranking metrics. Significantly outperforming conventional scoring methods, DeepRLI also shows marked superiority over recent machine learning-based approaches, including GenScore, PIGNet, and PLANET. These findings underscore the efficacy of DeepRLI as a multi-objective, physics-informed, contrast-optimized model. Its versatility and advanced capabilities position it as an integral tool for diverse computational tasks in drug design, encompassing affinity prediction, molecular docking, and virtual screening.

\subsubsection{Evaluation on CSAR-NRC HiQ Benchmark}

\begin{figure}[!hbtp]
  \includegraphics[width=\textwidth]{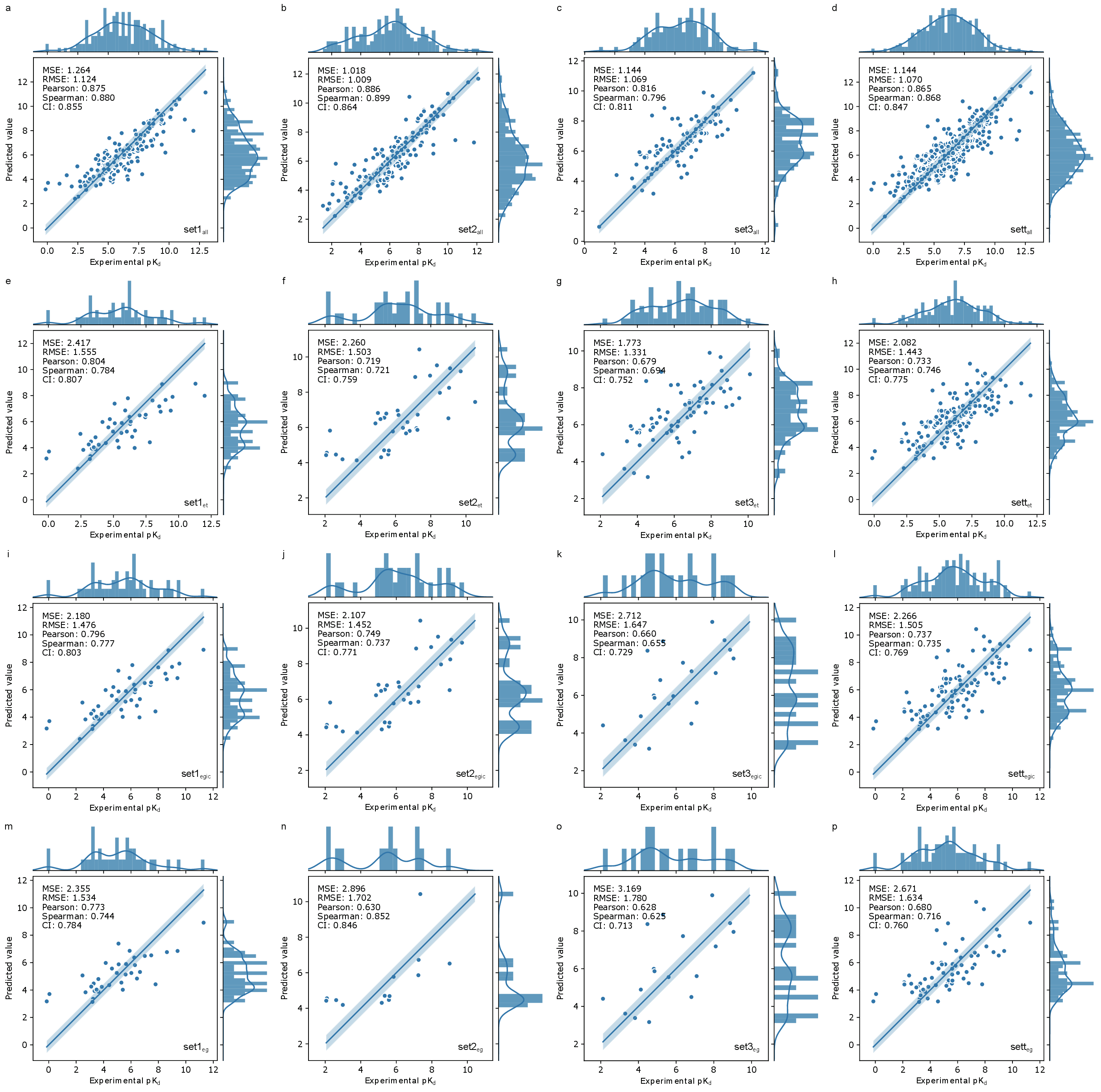}
  \caption{Scoring performance of DeepRLI on the CSAR-NRC HiQ benchmark. Scatter plots \textbf{a}-\textbf{p} show the correlation between predicted binding affinity values by DeepRLI and the actual experimental $pK_{\text{d}}$ values for various types of protein--ligand complex structure datasets. The scatter plots in each row involve datasets of types ``set1'', ``set2'', ``set3'', and ``sett'' (the union of the first three) from left to right, respectively. The scatter plots in each column involve datasets of types ``all'', ``et'' (exclude training set), ``egic'' (exclude general set, include core set), and ``eg'' (exclude general set) from top to bottom, respectively.}
  \label{fgr:csar_performance}
\end{figure}

Given the inherent limitations of analyzing performance based solely on a single benchmark due to its constrained dataset, our study extend the evaluation of DeepRLI to additional benchmarks beyond the confines of CASF-2016. A key part of this expanded analysis involve assessing the scoring power of DeepRLI on the CSAR-NRC HiQ benchmark, which is comprised of three distinct subsets, designated as $\text{set1}_{\text{all}}$ (comprising 176 complexes), $\text{set2}_{\text{all}}$ (167), and $\text{set3}_{\text{all}}$ (123). For comprehensive analysis, we aggregate these subsets into a collective set, referred to as $\text{sett}_{\text{all}}$, encompassing a total of 466 complexes. The performance of DeepRLI across these datasets is quantitatively evaluated, with results depicted in correlation scatter plots (Figures~\ref{fgr:csar_performance}-a, b, c, and d). Notably, the Pearson correlation coefficients between the predicted and experimental values are remarkably high, being 0.875, 0.886, 0.816, and 0.868, respectively. While these results might initially appear astonishing, further scrutiny reveals a critical issue of data leakage, wherein a portion of the training data is included in the test set, leading to an overestimation of scoring performance. Furthermore, the subpar scoring performance observed in $\text{set3}_{\text{all}}$ can be attributed primarily to the dataset's composition, which encompasses a substantial proportion of data entries annotated with imprecise $pK_{\text{d}}$ or $p\text{IC}_{50}$ values. The presence of these inaccurately measured experimental values is a significant factor contributing to the underestimation of the model's scoring capability.

\begin{table}[!hbtp]
  \caption{The scoring power of several representative scoring functions on the CSAR-NRC HiQ benchmark. Apart from DeepRLI, data for all other methods are from Shen \textit{et al.} \cite{shen2023ageneralized} The best results in each column are highlighted in bold}
  \label{tbl:performance_on_csar.scoring_power}
  \begin{tabular}{l cc cc}
    \toprule
    \multirow{2}{*}{Method} & \multicolumn{2}{c}{Scoring Power on $\text{sett}_{\text{egic}}$} & \multicolumn{2}{c}{Scoring Power on $\text{sett}_{\text{eg}}$} \\
    \cmidrule(lr){2-3} \cmidrule(lr){4-5}
    & $R_{\text{p}}$ & $\rho$ & $R_{\text{p}}$ & $\rho$ \\
    \midrule
    AutoDock4 \cite{huey2007asemiempirical} & 0.527 & 0.542 & 0.561 & 0.610 \\
    Vina & 0.306 & 0.589 & 0.282 & 0.543 \\
    Vinardo & 0.286 & 0.586 & 0.260 & 0.543 \\
    Glide SP & 0.126 & 0.571 & 0.115 & 0.551 \\
    Glide XP & 0.126 & 0.388 & 0.115 & 0.365 \\
    X-Score & 0.617 & 0.598 & 0.528 & 0.514 \\
    Pafnucy \cite{stepniewska-dziubinska2018development} & 0.610 & 0.625 & 0.583 & 0.605 \\
    GenScore & 0.713 & 0.697 & 0.678 & 0.674 \\
    DeepRLI & \textbf{0.737} & \textbf{0.735} & \textbf{0.680} & \textbf{0.716} \\
    \bottomrule
  \end{tabular}
\end{table}

To eliminate the impact of data leakage on performance evaluation, entries identical to those in the training set are meticulously excluded from the aforementioned datasets. This step leads to the generation of four reduced datasets, designated as $\text{set1}_{\text{et}}$ (50), $\text{set2}_{\text{et}}$ (36), $\text{set3}_{\text{et}}$ (75), and $\text{sett}_{\text{et}}$ (161), where ``et'' signifies the exclusion of the training set. The performance of DeepRLI on these reduced datasets is evaluated, with the results being graphically depicted through correlation scatter plots in Figures~\ref{fgr:csar_performance}-e, f, g, and h. Pearson correlation coefficients, measuring the congruence between predicted and experimental values, are found to be 0.804, 0.719, 0.679, and 0.733 for the respective datasets. These coefficients indicate a robust correlation across all datasets. Notably, DeepRLI's performance in these assessments underscores its commendable generalization capabilities, particularly in terms of scoring proficiency.

Additionally, for comparative analysis with other methods, especially the results of Shen \textit{et al.} \cite{shen2023ageneralized}, we further evaluate the scoring performance of DeepRLI on two types of datasets: one that excludes duplicates from the PDBbind general set but retains those belonging to the core set, and another that completely excludes duplicates from the general set. These datasets are respectively labeled ``egic'' (exclude general set, but include core set) and ``eg'' (exclude general set), namely $\text{set1}_{\text{egic}}$ (48), $\text{set2}_{\text{egic}}$ (33), $\text{set3}_{\text{egic}}$ (21), $\text{sett}_{\text{egic}}$ (102); $\text{set1}_{\text{eg}}$ (36), $\text{set2}_{\text{eg}}$ (13), $\text{set3}_{\text{eg}}$ (17), $\text{sett}_{\text{eg}}$ (66). The evaluation results of DeepRLI on these curated datasets are depicted in correlation scatter plots (Figure~\ref{fgr:csar_performance}-i, j, k, l, m, n, o and p). The Pearson correlation coefficients for the ``egic'' datasets are 0.796, 0.749, 0.660, and 0.737, respectively, while for the ``eg'' datasets, they are 0.773, 0.630, 0.628, and 0.680, respectively. These coefficients indicate a consistently robust correlation across all datasets. In Table~\ref{tbl:performance_on_csar.scoring_power}, we list the performance of various representative scoring functions on the $\text{sett}_{\text{egic}}$ and $\text{sett}_{\text{eg}}$ test sets. Notably, our DeepRLI model outperforms others in terms of both Pearson and Spearman correlation coefficients. This superior performance underscores the exceptional scoring accuracy and impressive generalization capability of our model, reinforcing its potential utility in computer-aided drug design for binding affinity prediction.

\subsubsection{Evaluation on Merck FEP Benchmark}

\begin{table}[!hbtp]
  \centering
  \caption{The ranking power, measured by Spearman correlation coefficient ($\rho$), of several representative scoring functions on the Merck FEP benchmark. Apart from DeepRLI, the data for all other models are from Shen \textit{et al.} \cite{shen2023ageneralized} The best result in each column is highlighted in bold}
  \label{tbl:performance_on_merck_fep.ranking_power}
  \resizebox{\columnwidth}{!}{
  \begin{tabular}{lccccccccc}
    \toprule
    Method & CDK8 & c-Met & Eg5 & HIF-2$\alpha$ & PFKFB3 & SHP-2 & SYK & TNKS2 & mean \\
    \midrule
    AutoDock4 & 0.629 & 0.324 & $-0.397$ & 0.376 & 0.530 & 0.609 & 0.544 & 0.558 & 0.397 \\
    Vina & \textbf{0.849} & $-0.257$ & $-0.520$ & \textbf{0.493} & 0.546 & 0.569 & 0.519 & 0.538 & 0.342 \\
    Vinardo & 0.782 & $-0.359$ & $-0.475$ & 0.371 & 0.515 & 0.490 & 0.379 & 0.305 & 0.251 \\
    Glide SP & 0.345 & 0.378 & $-0.111$ & 0.445 & 0.480 & 0.542 & $-0.006$ & 0.316 & 0.299 \\
    Glide XP & 0.617 & 0.165 & 0.017 & 0.410 & 0.513 & 0.490 & 0.124 & 0.582 & 0.365 \\
    X-Score & 0.406 & 0.531 & $-0.316$ & 0.224 & 0.430 & $-0.030$ & \textbf{0.689} & \textbf{0.669} & 0.325 \\
    MM-GBSA & 0.649 & 0.499 & $-0.002$ & 0.282 & 0.554 & 0.585 & 0.108 & 0.158 & 0.354 \\
    $\Delta_{\text{Lin\_F9}}$XGB & 0.826 & 0.077 & $-0.099$ & 0.480 & \textbf{0.603} & \textbf{0.640} & 0.103 & 0.458 & 0.386 \\
    Pafnucy & 0.406 & 0.531 & $-0.316$ & 0.224 & 0.430 & $-0.030$ & \textbf{0.689} & 0.669 & 0.325 \\
    GenScore & 0.675 & 0.677 & \textbf{0.275} & 0.437 & 0.571 & 0.338 & 0.144 & 0.578 & \textbf{0.462} \\
    DeepRLI & 0.513 & \textbf{0.745} & $-0.024$ & 0.459 & 0.577 & 0.639 & 0.441 & 0.331 & 0.460 \\
    \bottomrule
  \end{tabular}}
\end{table}

We further evaluate the ranking capability of DeepRLI on the Merck FEP benchmark. Originally, the Merck FEP benchmark is designed to assess the precision of various computational approaches in determining relative binding free energies based on fundamental physical principles. A notable characteristic of this dataset is the minimal variance among active small molecules targeting the same biomolecular target, presenting a significant challenge for scoring functions in accurately ranking these molecules. In our approach, we integrate DeepRLI scoring scores with docking scores to systematically rank small molecules across eight distinct targets within the dataset. The outcomes of this analysis are detailed in Table~\ref{tbl:performance_on_merck_fep.ranking_power}, which shows an average Spearman correlation coefficient of 0.460. While this ranking performance is moderate, it places DeepRLI amongst the leading methods in the field, highlighting its alignment with the current state-of-the-art. Most importantly, DeepRLI demonstrated exceptional performance in ranking molecules targeting the c-Met protein \cite{dieter2015identification}, achieving a Spearman correlation coefficient of 0.745, thereby outperforming all comparative methodologies. This result underscores the potential of our method in facilitating hit-to-lead and lead optimization processes, particularly for specific target proteins.

\subsubsection{Evaluation on LIT-PCBA Benchmark}

\begin{figure}[!hbt]
  \includegraphics[width=\textwidth]{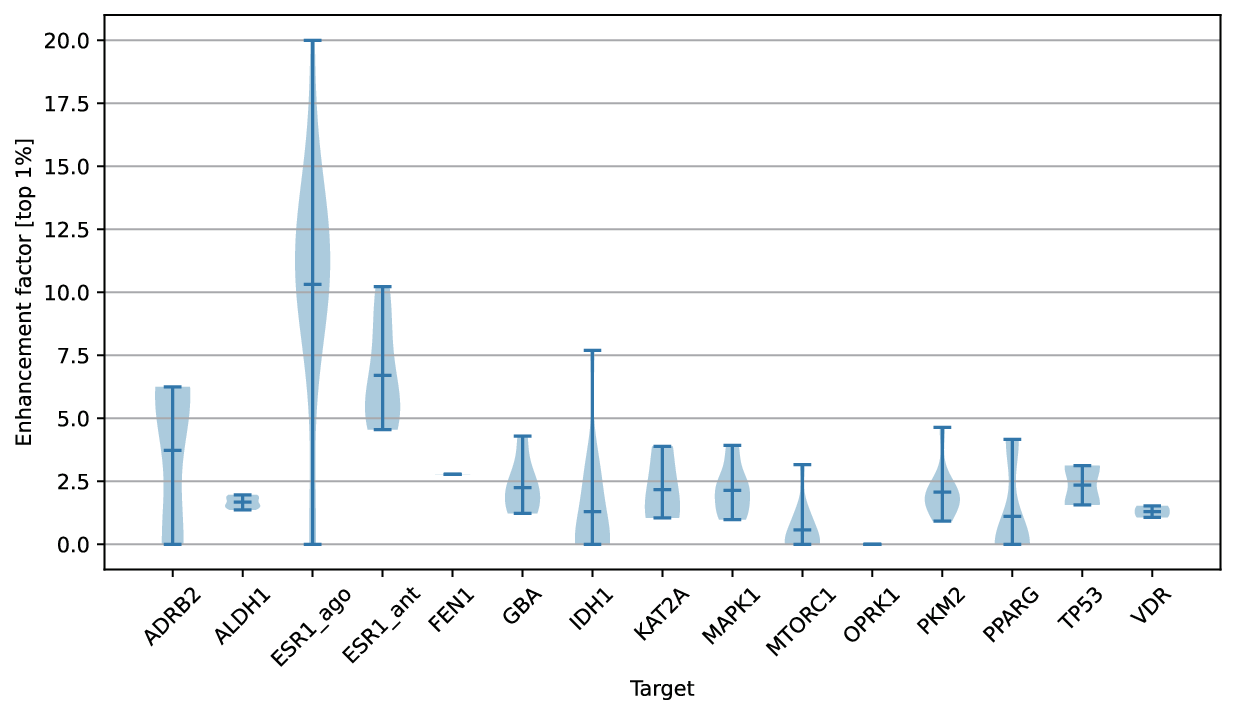}
  \caption{A violin plot showing the screening performance of DeepRLI on the LIT-PCBA benchmark. Each target has one or more PDB templates. And each section in the figure depicts the distribution of enhancement factors in the top 1\% measured by DeepRLI on different PDB templates of a target, with short horizontal lines marking the positions of the extremes and the mean.}
  \label{fgr:lit-pcba_performance}
\end{figure}

To further explore the screening capability of DeepRLI, we evaluated its performance on the well-crafted LIT-PCBA benchmark that mimics a real virtual screening scenario (with active and inactive data derived from experimental validations, and the distribution of chemical features of active and inactive molecules being similar, but with inactive molecules far outnumbering active ones). The screening results of DeepRLI for each target in the dataset are presented in Figure~\ref{fgr:lit-pcba_performance}, using the top 1\% enrichment factor as an indicator. It is noteworthy that the majority of the targets encompass several PDB templates. Variations in binding site conformations across these templates can exert differential impacts on our model's virtual screening efficacy. A detailed examination of Figure~\ref{fgr:lit-pcba_performance} reveals that for certain targets, namely ADRB2, ESR\_ago, ESR\_ant, and IDH1, there exists a pronounced variability in the top 1\% enrichment factor across different PDB templates, with disparities extending beyond a value of 5. In contrast, for other targets, while the disparities in outcomes across various PDB templates are relatively marginal, they are consistently minor. Overall, the best results of DeepRLI on each target are generally satisfactory, with an average top 1\% enrichment factor of 5.18, demonstrating basic screening proficiency.

\begin{table}[!hbt]
  \centering
  \caption{The screening power, measured by enhancement factor in the top 1\% ($\text{EF}_{1\%}$), of several representative scoring functions on the LIT-PCBA benchmark. The data for Vina, Lin\_F9, and $\Delta_{\text{vina}}\text{RF}_{20}$ are from Yang \textit{et al.} \cite{yang2022delta}, the data for Glide SP are from Shen \textit{et al.} \cite{shen2023ageneralized}, and for other methods except DeepRLI, the data are from their respective original publications. The best result in each row is highlighted in bold.}
  \label{tbl:performance_on_lit-pcba.screening_power}
  \resizebox{\columnwidth}{!}{
  \begin{tabular}{lcccccccc}
    \toprule
    Target & Vina & Glide SP & Lin\_F9 & $\Delta_{\text{vina}}\text{RF}_{20}$ & $\Delta_{\text{Lin\_F9}}$XGB & PLANET & GenScore & DeepRLI \\
    \midrule
    ADRB2 & 0.00 & 5.88 & 0.00 & 0.00 & 11.76 & 5.88 & \textbf{15.69} & 6.25 \\
    ALDH1 & 1.49 & 2.02 & 1.59 & 1.66 & \textbf{6.46} & 1.38 & 1.96 & 1.96 \\
    ESR1\_ago & 15.38 & 7.69 & 0.00 & 15.38 & 7.69 & 7.69 & 10.25 & \textbf{20.00} \\
    ESR1\_ant & 3.92 & 1.94 & 2.94 & 2.94 & 3.92 & 3.88 & 3.56 & \textbf{10.23} \\
    FEN1 & 0.54 & \textbf{7.32} & 1.90 & 0.81 & 2.17 & 5.15 & 6.05 & 2.78 \\
    GBA & 4.82 & 4.22 & 7.23 & 6.63 & \textbf{9.64} & 3.01 & 1.41 & 4.29 \\
    IDH1 & 0.00 & 0.00 & 2.56 & 0.00 & 5.13 & 2.56 & 5.13 & \textbf{7.69} \\
    KAT2A & 0.52 & 1.03 & 2.06 & 0.52 & \textbf{7.73} & 3.11 & 1.20 & 3.89 \\
    MAPK1 & 2.92 & 3.24 & 1.62 & 1.95 & 2.60 & 1.30 & \textbf{4.87} & 3.92 \\
    MTORC1 & 2.06 & 0.00 & 2.06 & 3.09 & 2.06 & 2.06 & 2.40 & \textbf{3.16} \\
    OPRK1 & 0.00 & 0.00 & 4.17 & 0.00 & \textbf{12.50} & 4.17 & 2.78 & 0.00 \\
    PKM2 & 1.65 & 2.75 & 0.73 & 2.93 & 2.56 & 1.83 & 1.47 & \textbf{4.64} \\
    PPARG & 7.41 & \textbf{21.96} & 3.70 & 11.11 & 7.41 & 3.66 & 20.74 & 4.17 \\
    TP53 & 0.00 & 2.50 & 2.53 & 0.00 & 1.27 & 2.50 & 0.00 & \textbf{3.12} \\
    VDR & 1.02 & 0.34 & 0.11 & 0.68 & 0.34 & 1.02 & 1.13 & \textbf{1.53} \\
    \midrule
    mean & 2.78 & 4.06 & 2.21 & 3.18 & \textbf{5.55} & 3.28 & 5.24 & 5.18 \\
    median & 1.49 & 2.50 & 2.06 & 1.66 & \textbf{5.13} & 3.01 & 2.78 & 3.92 \\
    max & 15.38 & \textbf{21.96} & 7.23 & 15.38 & 12.50 & 7.69 & 20.74 & 20.00 \\
    \midrule
    > 2 & 6 & 9 & 8 & 6 & \textbf{13} & 11 & 9 & 12 \\
    > 5 & 2 & 4 & 1 & 3 & \textbf{8} & 3 & 5 & 4 \\
    > 10 & 1 & 1 & 0 & 2 & 2 & 0 & \textbf{3} & 2 \\
    \bottomrule
  \end{tabular}}
\end{table}

In Table~\ref{tbl:performance_on_lit-pcba.screening_power}, we have listed the performance of some representative scoring functions on the LIT-PCBA benchmark. The results of other methods for each target are primarily based on a selected PDB template, so we also sampled a PDB template with the best result for comparison. As can be seen from the table, our DeepRLI model demonstrates a nice screening performance on all targets, ranking at the current advanced level, with an average $\text{EF}_{1\%}$ of 5.18, a median of 3.92, and a maximum of 20.00. A more detailed examination reveals that the DeepRLI model achieved an $\text{EF}_{1\%}$ of over 2 in 12 targets, surpassed 5 in 4 targets, and exceeded 10 in 2 targets. Compared to other scoring models, this is a fairly good outcome. It is noteworthy that, among 15 targets, DeepRLI's screening $\text{EF}_{1\%}$ is higher than other compared methods in 7 targets, indicating that our model can make reasonable predictions of active molecules for most targets, rather than performing exceptionally only on certain ones. Although our method does not demonstrate outstanding screening capabilities on CSAF-2016, it shows a performance close to the current advanced methods on the large-scale virtual screening benchmark LIT-PCBA, indicating that DeepRLI has superior generalization ability and can make reasonable screening inferences in external test sets.

\subsection{Interpretation}

Our model, which is based on the graph transformer layers, allows us to analyze its scoring decisions by extracting its attention weights. This model presents a natural advantage, as it predicts not only the affinity but also the interaction mode between the protein and the ligand. Given that the final prediction made by the DeepRLI model is closely related to the last layer of the graph transformer, we extract its attention weights, $w_{i j}^{k, L}$, and conduct a relevant analysis based on them.

In our approach, a graph typically consists of tens of thousands of edges, each with corresponding attention weights, making it challenging to display them all. Generally, our primary interest lies in the components involving both the protein and the ligand. By carefully examining these aspects, we can gain insight into which interactions play a more crucial role in the binding strength. Moreover, our model employs a multi-head attention mechanism within the graph transformer layers, comprising eight heads. To facilitate visualization, we compute the average of the attention weights across these eight heads.

\begin{figure}[!hbtp]
  \includegraphics[height=0.8\textheight]{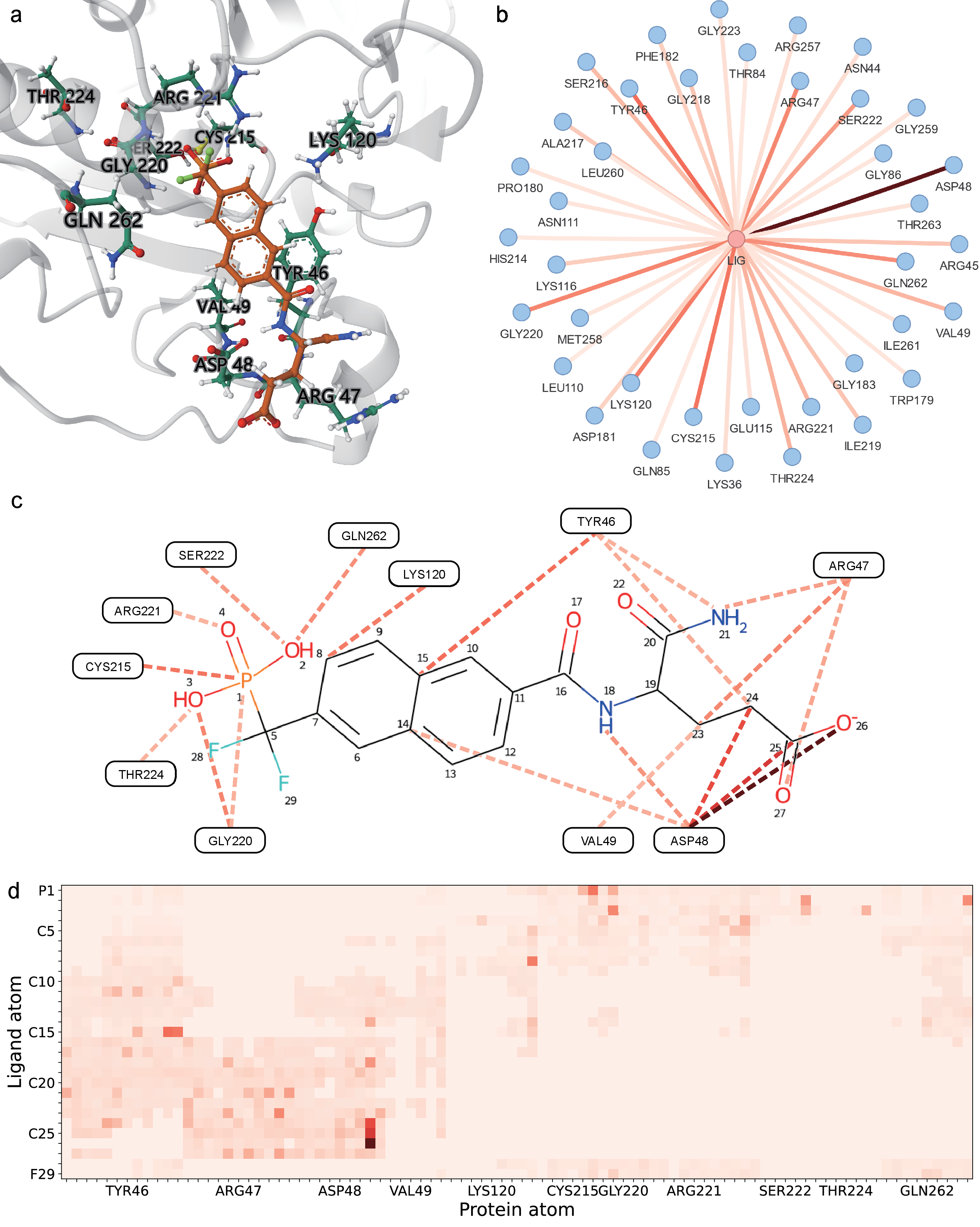}
  \caption{Visualization of interactions based on the attention weights from the final graph transformer layer of DeepRLI. Darker colors represent higher attention weights and more important interactions. The protein--ligand complex examined here is 1BZC. \textbf{a}, The 3D structure of the binding site of the protein--ligand complex, residues with higher attention weights are additionally shown in ball-and-stick representation. \textbf{b}, The graph displaying ligand--residue interaction connections. \textbf{c}, The graph depicting [ligand atom]--residue interaction connections. \textbf{d}, The heatmap illustrating [ligand atom]--[residue atom] interaction connections; For clarity, only part of interaction connections with a larger weight are shown for the latter three.}
  \label{fgr:deeprli.interpretation.attention}
\end{figure}

In this manner, we can obtain not only the averaged attention weights of protein--ligand atom pairs but also the attention strength of a specific residue towards a ligand atom and even the importance of each residue for the entire ligand by aggregating a set of weights. The aggregation method employed here is ``max'', where the maximum weight is used for visualization purposes. We sequentially map numeric values to colors for visualizing attention weights, with Figure~\ref{fgr:deeprli.interpretation.attention} providing an example of the system of human PTP1B catalytic domain complexed with TPI (PDB ID: 1BZC) \cite{groves1998structural}. It can be seen that relatively large attention weights appear between the ligand and TYR46, ARG47, ASP48, etc., indicating important interactions between them. To validate these findings, we employed PLIP (Protein--Ligand Interaction Profiler) as an analytical tool \cite{salentin2015plip,adasme2021plip}. Our analysis revealed that TYR46 engages in $\pi$-stacking and hydrophobic interactions with the ligand, while ARG47 and ASP48 form hydrogen bonds. The comprehensive PLIP results are documented in Tables~\ref{tbl:plip_1bzc_hydrophobic_interactions}, \ref{tbl:plip_1bzc_hydrogen_bonds}, and \ref{tbl:plip_1bzc_pi_stackings}. The consistency between our model's predictions and PLIP's analysis substantiates our model's capability to accurately identify key interactions based on the atoms' surrounding environment, thereby elucidating a portion of its robust inference ability.

\begin{table}[!hbtp]
  \caption{Hydrophobic interactions in the 1BZC complex analyzed by PLIP}
  \label{tbl:plip_1bzc_hydrophobic_interactions}
  \begin{tabular}{cccc}
    \hline
    Residue & Distance & Ligand Atom & Protein Atom \\
    \hline
    TYR46 & 3.94 & 2438 & 377 \\
    VAL49 & 3.91 & 2442 & 405 \\
    PHE182 & 3.35 & 2436 & 1527 \\
    ALA217 & 3.45 & 2435 & 1786 \\
    \hline
  \end{tabular}
\end{table}

\begin{table}[!hbtp]
  \caption{Hydrogen bonds in the 1BZC complex analyzed by PLIP}
  \label{tbl:plip_1bzc_hydrogen_bonds}
  \resizebox{\columnwidth}{!}{
  \begin{tabular}{ccccccc}
    \hline
    Residue & Distance H-A & Distance D-A & Donor Angle & Donor Atom & Acceptor Atom \\
    \hline
    ARG47 & 3.27 & 4.03 & 135.35 & 389 [Ng+] & 2456 [O.co2] \\
    ARG47 & 2.27 & 3.19 & 155.66 & 380 [Nam] & 2451 [O2] \\
    ASP48 & 2.94 & 3.90 & 165.69 & 2447 [Nam] & 398 [O3] \\
    SER216 & 2.02 & 2.73 & 127.06 & 1776 [Nam] & 2431 [O3] \\
    ALA217 & 1.87 & 2.81 & 159.54 & 1782 [Nam] & 2431 [O3] \\
    GLY218 & 2.80 & 3.30 & 111.55 & 1787 [Nam] & 2429 [O2] \\
    ILE219 & 2.17 & 3.10 & 156.43 & 1791 [Nam] & 2429 [O2] \\
    GLY220 & 1.96 & 2.94 & 172.50 & 1799 [Nam] & 2429 [O2] \\
    ARG221 & 1.84 & 2.81 & 169.06 & 1803 [Nam] & 2430 [O3] \\
    ARG221 & 1.79 & 2.72 & 155.79 & 1813 [Ng+] & 2431 [O3] \\
    \hline
  \end{tabular}}
\end{table}

\begin{table}[!hbtp]
  \caption{$\pi$-stackings in the 1BZC complex analyzed by PLIP}
  \label{tbl:plip_1bzc_pi_stackings}
  \resizebox{\columnwidth}{!}{
  \begin{tabular}{ccccccc}
    \hline
    Residue & Distance & Angle & Offset & Stacking Type & Ligand Atoms \\
    \hline
    TYR46 & 3.92 & 21.79 & 0.63 & Parallel & 2439, 2440, 2441, 2442, 2443, 2444 \\
    \hline
  \end{tabular}}
\end{table}

\section{Conclusions}

In this study, we propose DeepRLI, a novel multi-objective deep learning framework designed for the universal protein--ligand interaction prediction. DeepRLI employs a fully connected graph as its input, effectively preserving the molecular topology and spatial structures. And this framework uses an improved graph transformer layer, combined with cosine constraints, which facilitates robust feature embedding. Central to DeepRLI's architecture are three distinct downstream networks, each dedicated to a specific predictive task: scoring, docking, and screening. The scoring readout network aims to accurately predict the binding free energy of crystal structures using a series of basic fully connected layers. Meanwhile, the docking and screening readout networks focus on identifying the optimal binding poses and enriching active small molecules, respectively. A key characteristic in these networks is the integration of physical information blocks, designed to improve the model's inference capability, especially for protein--ligand conformations with loose bindings that deviate from typical crystal structures. To further enhance the model's generalization ability, we incorporated data augmentation techniques, including re-docking and cross-docking procedures to generate more data, complemented by contrastive learning strategies. This combination enhances DeepRLI's applicability across diverse datasets and scenarios.

DeepRLI's efficacy is rigorously tested on several established benchmarks. Its performance is evaluated across a range of tasks - scoring, ranking, docking, and screening - on the CASF-2016, CSAR-NRC HiQ, Merck FEP, and LIT-PCBA benchmarks. The results consistently demonstrate DeepRLI's superior inferential abilities in all tested domains, underscoring its versatility as a predictive tool for protein--ligand interactions.

Furthermore, the implementation of a graph transformer layer as the primary feature embedding module in DeepRLI offers notable interpretability advantages. The model assigns greater attention weights to edges that signify crucial interaction patterns, providing insights into the underlying molecular interactions.

In conclusion, the DeepRLI framework can effectively provide useful guidance for structure-based drug design. Its innovative approach and proven efficacy in predicting protein--ligand interactions position it as a powerful and versatile tool in the field of drug discovery.

\section*{Author Contributions}

H. L. conceived the basic idea. H. L. designed and implemented the deep learning model and performed the model training, evaluation and interpretation. H. L., S. W. and J. Z. participated in the processing of training data and test data. H. L., S. W. and Y. L. discussed some details of the concepts and methods. J. P. and L. L. supervised the project. H. L. wrote the manuscript.

\section*{Funding}

This work was supported in part by the National Natural Science Foundation of China (Grants 22033001, 21673010 and 21633001).

\begin{acknowledgement}

We express our gratitude to all members of Luhua Lai's group for their valuable suggestions and insights. We also acknowledge the support of computing resources provided by the high-performance computing platform at the Peking--Tsinghua Center for Life Sciences, Peking University.

\end{acknowledgement}

\begin{suppinfo}

For code and other supplementary information, please refer to \url{https://github.com/fairydance/DeepRLI}.

\end{suppinfo}

\bibliography{deeprli-manuscript}

\end{document}